\pdfoutput=1
\documentclass[twocolumn,showkeys,showpacs,preprintnumbers,prd,superscriptaddress,nofootinbib]{revtex4-1}
\usepackage{graphicx}
\usepackage{epsf}
\usepackage{bm}
\usepackage{amsmath}
\usepackage{amsfonts}
\usepackage{amssymb}
\usepackage{epstopdf}
\usepackage{natbib}
\usepackage{hyperref}
\usepackage{color}
\usepackage{verbatim}
\usepackage{multirow}
\usepackage{bm}
\usepackage{hyperref}
\usepackage{url}
\usepackage{orcidlink}
\usepackage{wrapfig}
\usepackage{nicefrac, xfrac}
\usepackage{enumitem}
\usepackage[normalem]{ulem}

\usepackage{hyperref}

\makeatletter\let\expandableinput\@@input\makeatother

\begin{document}

\title{Matter–antimatter asymmetry in a rotating universe: Dirac spinors in axisymmetric Bianchi IX cosmology}

\author{Tatevik Vardanyan}
\email{tatevik@thp.uni-koeln.de}
\affiliation{University of Cologne, Faculty of Mathematics and Natural Sciences, Institute for Theoretical Physics, Cologne, Germany}

\begin{abstract}

The standard $\Lambda$ CDM model, based on a highly symmetric FLRW geometry, successfully explains many observations but faces unresolved issues, motivating the exploration of alternative cosmological frameworks. We investigate the influence of spacetime geometry on the matter–antimatter asymmetry of the Universe within the Bianchi IX cosmological model. Motivated by early-universe dynamics, potential contributions to cosmological angular momentum, and the axisymmetric Bianchi IX model’s relevance to certain CMB anomalies, we formulate the Dirac field in this background. Starting with a Lagrangian formalism, we derive the Dirac equation and develop the harmonic analysis of spinor fields, extending previous treatments in the Mixmaster universe. We solve the Dirac equations for non-rotating and rotating axisymmetric Bianchi IX spacetimes using a fixed-background approximation. We find that spatial anisotropy induces spin-dependent energy splittings, while global rotation produces particle–antiparticle asymmetries in the energy spectra, effects absent in FLRW models. These results demonstrate that spacetime geometry alone can imprint nontrivial structure on particle spectra, suggesting that geometric effects may contribute to the matter–antimatter asymmetry.

\end{abstract}



\maketitle

\section{Introduction} \label{sec:introduction}

The standard model of cosmology, the $\Lambda$CDM model, has proven highly successful in accounting for a wide range of observations, including measurements of the Cosmic Microwave Background (CMB) from WMAP \cite{paper:Wmap} and Planck \cite{paper:Planck}, as well as large-scale structure data from the Sloan Digital Sky Survey (SDSS) \cite{paper:Ad}, among others. Nevertheless, it faces persistent puzzles—including the nature of dark matter and dark energy, matter–antimatter asymmetry, the Hubble tension, and anomalies in the CMB \cite{paper:Greek}—as well as small-scale challenges in the cold dark matter paradigm \cite{paper:smallscalep}. These tensions suggest that $\Lambda$CDM may be incomplete and point to the need for a revision or extension of its underlying assumptions, particularly the validity of the highly symmetric Friedmann-Lemaître-Robertson-Walker (FLRW) geometry~\cite{Aluri:2022hzs}. 
Relaxing the assumption of isotropy and considering more general cosmological models may therefore provide new insights into unresolved puzzles.

Motivated by this perspective, we investigate whether the matter–antimatter asymmetry can originate in the geometric structure of spacetime itself, rather than requiring physics beyond the Standard Model. Specifically, we consider the Bianchi IX cosmological model, the choice of which is motivated by its relevance to early-universe dynamics, especially in connection with the BKL conjecture \cite{paper:BKL1,paper:BKL2} and its potential role in angular momentum generation within a cosmological context \cite{paper:Li}. Any realistic cosmological model must remain consistent with CMB observations; accordingly, if global rotation exists, its angular velocity is constrained by CMB limits to approximately $\omega \sim 10^{-12}$ rad $\mathrm{yr}^{-1}$ \cite{book:CW}.

The ultimate goal in studies of matter–antimatter asymmetry is to address the baryon asymmetry within Quantum Chromodynamics (QCD) in a Bianchi IX cosmological background. However, as a natural first step, this work focuses on the simpler case of the Dirac fermion field, with the aim of exploring how the geometry of the Bianchi IX background influences the energy spectra of particles and antiparticles, potentially leading to an asymmetry relevant to the matter–antimatter imbalance. Earlier, Gibbons \cite{paper:Gib1, paper:Gib3} demonstrated that anisotropic cosmological backgrounds can produce spectral asymmetries between Weyl neutrinos and antineutrinos, thereby suggesting that geometric anisotropy itself may contribute to matter–antimatter asymmetry. This conceptual insight motivates the present investigation in the context of Dirac fields in a Bianchi IX background.

We develop the Lagrangian formulation of a Dirac spinor field in a general Bianchi IX cosmological background, extending earlier studies of homogeneous Dirac fields \cite{paper:ORyan, paper:Damour}, and derive the corresponding Dirac equation. To make the problem analytically tractable, we focus on the axisymmetric subclass of Bianchi IX geometries, for which the Dirac equation can be solved explicitly. This choice captures essential features of the general Bianchi IX class and is further motivated by potential connections to observed CMB anomalies \cite{paper:Axisym}. While the fixed-background assumption is a simplification, it provides a controlled setting to isolate the effects of anisotropy and rotation on particle and antiparticle spectra. A fully realistic treatment, including time-dependent backgrounds and adiabatic or semiclassical refinements based on the WKB approximation \cite{book:Birrell}, is left for future work. Within this framework, we examine how geometric anisotropies influence particle and antiparticle energy spectra in both non-rotating and rotating cases, revealing novel features absent in isotropic cosmologies and highlighting the potential role of anisotropic geometry in matter–antimatter asymmetry\footnote{This paper is based on part of the author’s Ph.D. thesis  submitted to the University of Cologne \cite{Vardanyan2025thesis}.}.

We develop the harmonic analysis of spinor fields in the Bianchi IX background using the group-theoretic structure of its spatial symmetries. Earlier work by Dowker and Pettengill \cite{paper:Dowker}, building on Hu’s scalar-field analysis \cite{paper:HuI}, developed the mode decomposition for spinors in the Mixmaster universe and reduced it to the eigenvalue problem of an ``ideal'' asymmetric top Hamiltonian. In the present work, we extend this approach to the general Bianchi IX geometry, which admits global rotation.

The paper is organized as follows. In Sec.~\ref{puzzle}, we present the conceptual foundations of this work by reviewing key cosmological puzzles and addressing them from a geometric perspective, motivating the choice of the Bianchi IX model. In Sec.~\ref{Diraceq}, we introduce the Bianchi IX geometry, develop the Lagrangian formulation of a Dirac spinor field in this background and derive the corresponding Dirac equation. Sec.~\ref{HA} presents the harmonic analysis of spinor fields in this geometry. In Sec.~\ref{sol}, we solve the Dirac equation in axisymmetric Bianchi IX cosmology and analyze the resulting particle and antiparticle spectra. The conclusions and outlook are presented in Sec.~\ref{con}. Appendix~\ref{irr} reviews the relevant unitary irreducible representations of the group $\mathrm{SU(2)}$, which underlie the mode decomposition of spinor fields in Bianchi IX backgrounds. In Appendix~\ref{Atop}, we construct the eigenstates of the ideal asymmetric top Hamiltonian.

\section{Conceptual foundations: cosmological puzzles and geometric motivation}  \label{puzzle}

To motivate the geometric approach pursued in this work, this section reviews several outstanding puzzles in cosmology, such as the CMB anomalies, dark matter, and matter–antimatter asymmetry, and discusses how Bianchi IX geometry may offer a more natural setting for addressing these issues.

\subsection{Puzzles of \texorpdfstring{$\Lambda$CDM}{LambdaCDM} cosmology}

\subsubsection{CMB anomalies}

Measurements of the CMB by missions such as COBE \cite{paper:Cobe}, WMAP \cite{paper:Wmap}, and Planck \cite{paper:Planck}, with increasing precision, have revealed unexpected large-scale features. These anomalies challenge the expected statistical isotropy and Gaussianity of CMB temperature fluctuations—underlying assumptions of the $\Lambda$CDM model, with statistical isotropy following from the cosmological principle and Gaussianity arising from inflation \cite{paper:Dom}.

The following key anomalies have been observed: the \textit{lack of large-angle temperature correlations}, with the correlation function nearly vanishing for $\theta \gtrsim 60^\circ$; the \textit{quadrupole–octupole alignment}, referring to the unexpected alignment and near coplanarity of the quadrupole ($\ell=2$) and octupole ($\ell=3$) multipoles, whose common plane lies roughly perpendicular to the Ecliptic and aligns with the CMB dipole—suggesting a possible preferred direction; a \textit{hemispherical asymmetry} in temperature fluctuations; a \textit{parity asymmetry} between odd and even multipoles at the largest angular scales; and a large \textit{cold spot} in the southern CMB sky. For comprehensive discussions of these anomalies, see, e.g., \cite{paper:Greek}, \cite{paper:CMBa}, \cite{paper:Dom}.

These CMB anomalies present significant challenges to the predictions of the standard $\Lambda$CDM model. Although some have argued they may result from statistical fluctuations \cite{paper:stat}, the simultaneous occurrence of multiple independent anomalies makes this explanation increasingly unlikely \cite{paper:CMBa}. Moreover, neither large-scale systematics nor foreground contamination adequately account for these features, pointing instead to a possible cosmological origin. Leading theoretical candidates include non-trivial cosmic topology and anisotropic geometries, which offer promising frameworks to explain these persistent anomalies \cite{paper:ab}.

\subsubsection{Dark matter}

Compelling evidence for the existence of dark matter comes from various gravitational effects\footnote{Alternative gravity theories, most notably MOND, have also been explored as attempts to explain the effects attributed to dark matter without introducing additional matter, but they generally fail to account for all observed phenomena \cite{paper:Fam}.}, including galaxy rotation curves, gravitational lensing, galaxy cluster dynamics, cluster collisions, and the CMB data (see, e.g., \cite{notes:Blas, paper:Kat}). These observations suggest that dark matter possesses several defining properties: extremely weak interaction with Standard Model particles (primarily via gravity), near collisionlessness, coldness (i.e., non-relativistic), stability, and a non-baryonic nature. The latter is supported by both CMB measurements and Big Bang Nucleosynthesis (BBN) estimates, which tightly constrain the baryonic matter fraction to about 5\% of the total energy density \cite{paper:Lars}. However, no Standard Model particle exhibits these properties, motivating a wide array of beyond-Standard Model candidates (see, e.g., \cite{paper:Silk}), although direct evidence for any of these proposed particles remains absent. 

Despite its success on large scales, the Cold Dark Matter (CDM) model faces persistent challenges at smaller scales, particularly those below approximately 1 Mpc, including the \textit{Cusp/Core}, \textit{Missing Satellites}, and \textit{Too Big to Fail} problems, as well as the \textit{angular momentum catastrophe} \cite{paper:smallscalep}, \cite{paper:Bosch}. Various proposed resolutions, including the Warm Dark Matter (WDM) model \cite{paper:Matt}, address some of the small-scale challenges of the CDM paradigm, yet none provides a conclusive solution. For a more detailed discussion of these issues and potential approaches, see \cite{paper:smallscalep}, \cite{paper:ssp}, \cite{paper:Fuzzy}, and references therein.

\subsubsection{Matter-antimatter asymmetry} \label{m-as}

The Universe is overwhelmingly composed of matter, with almost complete absence of antimatter. However, Quantum Field Theory (QFT)—the foundational framework of the Standard Model—predicts that pair creation and annihilation processes are symmetric with respect to particles and antiparticles. As a result, the observed matter–antimatter asymmetry remains an unresolved issue in modern cosmology\footnote{Some theories suggest the Universe may contain separate regions of matter and antimatter or a homogeneous mixture of both; however, no $\gamma$-ray signals from annihilation have been observed to support this \cite{paper:SS}, \cite{paper:antiU}.}.

The study of matter–antimatter asymmetry focuses primarily on baryon asymmetry, as it is directly observable and constitutes the main component of the matter density of the Universe aside from dark matter and dark energy. Leptons, such as neutrinos and electrons, may also exhibit asymmetries, though these are more challenging to measure directly. The electron asymmetry is inferred indirectly from the baryon asymmetry, as the charge neutrality of the Universe requires that the number of electrons matches the number of protons \cite{book:Rub}.

The baryon asymmetry of the Universe (BAU) is quantified by the baryon-to-photon ratio, $\eta = N_B / N_\gamma$, measured independently from CMB temperature fluctuations and light element abundances predicted by BBN \cite{paper:maas}. Both methods yield consistent values, $\eta \sim 10^{-10}$, implying that today’s pronounced matter-antimatter imbalance arose from a tiny excess in the early Universe \cite{paper:etaBBN, paper:etaCMB}. The generation of baryon asymmetry requires the three Sakharov conditions to be satisfied \cite{paper:Sakharov}: violation of baryon number ($B$), violation of C and CP symmetries, and departure from thermal equilibrium. While the Standard Model satisfies these criteria, these effects are too weak to explain the observed baryon asymmetry, making successful baryogenesis unlikely within the Standard Model alone \cite{paper:maas}, \cite{paper:Kuzmin}.

Consequently, various extensions of the Standard Model and beyond have been proposed to explain the observed baryon asymmetry by introducing additional sources of CP and baryon number violation. Among these are Electroweak Baryogenesis models, such as those with an extended Higgs sector or the Minimal Supersymmetric Standard Model (MSSM), as well as other notable frameworks including Grand Unified Theory (GUT) Baryogenesis, the Affleck-Dine mechanism in supersymmetric theories, Leptogenesis, and gravitational baryogenesis, which involves a CP-violating interaction between the Ricci scalar and the baryon current in an expanding Universe; see \cite{paper:GB}. While these scenarios remain promising, they continue to face significant unresolved theoretical and observational challenges  (see, e.g., \cite{paper:Dine, paper:Rub, paper:Trod} for detailed reviews).

\subsection{Geometric perspective on cosmological puzzles}

Given the challenges outlined, a natural question arises: are the foundational assumptions of the FLRW geometry in the $\Lambda$CDM model correct? Could the simplicity of such a highly symmetric framework be contributing to these unresolved issues? Thus, a promising approach to resolving these cosmological puzzles is to relax the assumption of isotropy, as suggested by the CMB anomalies, and to explore more general cosmological models based on more complex geometries, which can then be tested against observational data.

Such an assumption may also offer fresh perspectives on cosmological puzzles. Within the $\Lambda$CDM model, dark matter plays a central role in the growth of initial perturbations, i.e., seeds for the formation of large-scale structures, as sufficient gravitational attraction would not be possible without it \cite{book:Rubakov}. In homogeneous anisotropic cosmological models, the evolution of these perturbations differs from that in FLRW geometries (see, e.g., \cite{book:EMM} and references therein), thereby influencing the process of structure formation. As a result, certain phenomena commonly attributed to dark matter may instead be explained by anisotropic gravitational effects. However, as previously discussed, some key observations supporting the existence of dark matter, such as the Bullet Cluster, would be difficult to account for through anisotropy alone, without introducing actual dark matter.

In addressing the matter–antimatter asymmetry, it is important to note that, in contrast to the FLRW framework, where the only gravitational influence on particle physics compared to Minkowski spacetime is the isotropic expansion of the universe leading to particle redshift \cite{book:Grib}, anisotropic cosmological models introduce more intricate geometric effects as shown by Gibbons in \cite{paper:Gib1, paper:Gib3}. In suitably chosen anisotropic backgrounds, such effects may give rise to asymmetries between particles and antiparticles, potentially offering an explanation for the observed asymmetry without invoking mechanisms beyond the Standard Model.

Let us also point out that, since electric charge is conserved in general relativity, particles and antiparticles are expected to be produced in equal numbers. However, if spatial anisotropy induces a fundamental asymmetry, for example a suppression of antiparticles, matter and antimatter could coexist without complete annihilation. In this scenario, the antiparticle sector, together with anisotropic gravitational effects, might fully account for the phenomena attributed to dark matter, eliminating the need to invoke additional exotic matter. While this idea remains speculative, it offers an intriguing avenue for further investigation. A definitive conclusion requires detailed studies of Quantum Electrodynamics (QED) in a specific anisotropic cosmological model, a task reserved for future investigation.

In this work, we adopt the Bianchi IX cosmological model, which describes a closed, spatially homogeneous, and anisotropic universe, which in general admits global rotation. Our motivation for this choice stems primarily from the model’s relevance to near-singularity dynamics and its capacity to support angular momentum generation in cosmological context, both of which are briefly reviewed in the following subsections.

\subsection{Motivation for a general Bianchi IX universe}  \label{Mot}

\subsubsection{The BKL conjecture}

As the universe approaches the Big Bang singularity, which is a generic feature of Einstein’s field equations \cite{book:LandauF}, the isotropy of FLRW models cannot be maintained \cite{paper:Lif}. Instead, as predicted by the BKL conjecture, the dynamics become highly chaotic and oscillatory near the singularity, dominated by time-dependent factors rather than spatial variations \cite{paper:BKL1, paper:BKL2}. As a result, different spatial points effectively decouple, and the behavior of the universe resembles that of the general Bianchi IX model\footnote{The dynamics of the diagonal Bianchi IX model were independently studied by Misner \cite{paper:Misner69} in search of a solution to the horizon problem. These results were later generalized to symmetric and general Bianchi IX models by Ryan \cite{paper:Ryan1, paper:Ryan2}.}. Therefore, the Bianchi IX model provides a more realistic and reasonable framework for investigation, assuming that isotropization occurs under appropriate conditions and that the geometry evolves toward the FLRW model observed today.

Possible mechanisms for isotropization are the anisotropy damping caused by cosmological particle creation \cite{paper:HuP} and the inflationary phase. A general discussion on the possibility of inflation in anisotropic universes, and whether it leads to isotropization, is available in \cite{book:EMM} and references therein. In particular, \cite{paper:Is} investigates the isotropization of Bianchi IX models driven by a scalar field with an exponential potential $e^{k\varphi}$. It has been shown that for $k<\sqrt{2}$, a set of continuously expanding Bianchi IX models models evolves toward isotropy. Conversely, for $k > \sqrt{2}$, initially expanding Bianchi IX models fail to approach a continuously expanding isotropic state. 

It is important to note that the claim that inflation can evolve the universe from generic initial conditions toward a homogeneous and isotropic state depends on the validity of the ``cosmic no-hair conjecture'', proposed by Gibbons and Hawking \cite{paper:GH} and Hawking and Moss \cite{paper:HM}. Wald \cite{paper:Wald} analyzed this conjecture within homogeneous cosmologies, demonstrating that all initially expanding Bianchi models except type IX approach the de Sitter solution on a rapid exponential timescale. For Bianchi IX, a similar behavior occurs provided the cosmological constant is sufficiently large, supporting the applicability of the conjecture in this case.

\subsubsection{Global rotation and angular momentum generation}

The origin of galactic angular momentum remains a fundamental open problem in astrophysics and cosmology. Proposed explanations include primordial turbulence, tidal torques, and global rotation (see e.g., \cite{paper:Li}, \cite{book:Peebles} for further details). While primordial turbulence from early density fluctuations may have contributed, it likely could not have persisted long enough against dissipation processes to generate significant angular momentum. Furthermore, the tidal torque theory, which attributes the origin of galactic angular momentum to tidal forces exerted by neighboring proto-galaxies. According to this theory, angular momentum is generated through gravitational interactions between collapsing regions and their surrounding matter. However, its applicability is limited to the linear regime, where density perturbations remain small. Moreover, the theory has difficulty accounting for the observed relation $J\propto M^{5/3}$ commonly found in galaxies. These limitations make the hypothesis of global rotation in the early universe an intriguing alternative. 

Initially proposed by Gamow \cite{paper:Gamow}, Gödel \cite{paper:Godel1}, and Collins and Hawking \cite{paper:HawkC}, global rotation could naturally induce the rotation of galaxies during their formation through the Coriolis force within galactic frames. Subsequently, Li \cite{paper:Li} conducted a more detailed investigation into how global rotation might affect angular momentum, structure formation, and the evolution of cosmic objects. The influence of global rotation has also been further explored in works such as \cite{paper:Ob}, \cite{paper:Godl2}, \cite{paper:Godl11}, and others.

Furthermore, observations reveal angular momentum on unexpectedly large scales: galaxy filaments exhibit rotation and maintain coherent spin when stacked \cite{paper:Wang}, and some galaxy clusters also show rotational motion \cite{paper:rotcl}. The clusters could gain angular momentum from off-axis mergers or global cosmic rotation. However, studies have not found strong evidence of recent mergers in the clusters suspected to be rotating. Therefore, the global rotation could offer a plausible explanation for their angular momentum.

Having established the conceptual foundations and motivation for the geometric approach, we now turn to the mathematical formulation by introducing the Dirac field in the Bianchi IX background.

\section{Dirac Spinor Field in Bianchi IX Cosmology} \label{Diraceq}

\subsection{The line element}

Spatially homogeneous Bianchi models admit a three-dimensional Lie group of isometries \(G_3 \) acting transitively on spacelike hypersurfaces and are classified according to the structure of the associated Lie algebra (see e.g., \cite{book:EMM}). In such spacetimes, the ADM line element takes the form\footnote{The ADM decomposition is used to keep the setup general and compatible with Hamiltonian formulation \cite{book:Kiefer}, though here we focus on the field equations. }
\begin{equation}
    ds^2 = -N^2 dt^2 + h_{ij} \left(N^i dt + \sigma^i\right)\left(N^j dt + \sigma^j\right),
    \label{hijsigma}
\end{equation}
where \( h_{ij} \) is the spatial metric, \( N \) is the lapse function, \( N^i \) is the shift vector, and \( \sigma^i \) with \(i = 1,\, 2,\, 3\) are the one-forms dual to the invariant basis \(\{ e_i \}\) introduced below.

The symmetry group of the Bianchi IX model is the rotation group \(\mathrm{SO}(3)\), with structure constants \(C^k_{ij} = -\varepsilon_{ijk}\). The topology of this model is that of a 3-sphere, \(S^3\). For Bianchi IX, the automorphism and special automorphism groups coincide with the symmetry group of the homogeneous space, allowing the spatial metric to be diagonalized via the rotation group \cite{paper:Jantzen}. Thus, the three-metric \(h_{ij}\) for the Bianchi IX model can be expressed as  
\begin{equation}
   h_{ij} = \Bar{h}_{kl} {R^k}_i {R^l}_j,
   \label{B3m}
\end{equation}
where \(R \in \mathrm{SO}(3)\) is a rotation matrix parameterized by Euler angles \(\{\phi, \theta, \psi\}\) in $zyz$-convention, with ranges \(0 \leq \phi \leq 2\pi\), \(0 \leq \theta \leq \pi\), and \(0 \leq \psi \leq 2\pi\), and can be written as the product of three successive rotations as follows:
\begin{equation}
    R = R_z(\psi) R_y(\theta) R_z(\phi),
    \label{RRR}
\end{equation}
where
\begin{align}
R_z(\phi)=\left(
\begin{matrix}
   \cos{\phi} & \sin{\phi} & 0\\
   -\sin{\phi} & \cos{\phi} & 0\\
   0 & 0 & 1
\end{matrix}
\right),
\label{Rz1}
\end{align}
\begin{align}
R_y(\theta)=\left(
\begin{matrix}
   \cos{\theta} & 0 & -\sin{\theta}\\
   0 & 1 & 0\\
   \sin{\theta} & 0 & \cos{\theta}
\end{matrix}
\right),
\label{Ry}
\end{align}
\begin{align}
R_z(\psi)=\left(
\begin{matrix}
   \cos{\psi} & \sin{\psi} & 0\\
   -\sin{\psi} & \cos{\psi} & 0\\
   0 & 0 & 1
\end{matrix}
\right).
\label{Rz2}
\end{align}
The diagonal three-metric $\Bar{h}_{kl}$, expressed in terms of the Misner variables $\alpha$, $\beta_+$, and $\beta_-$ \cite{paper:Misner69}, \cite{paper:Misner68}, takes the form
\begin{equation}
    \Bar{h}_{kl} = \left[ \textnormal{diag} \left( e^{\tilde{\beta}_1 }, e^{\tilde{\beta}_2}, e^{\tilde{\beta}_3}\right) \right]_{kl} \quad \textnormal{with} \quad \Tilde{\beta}_k = 2(\beta_k + \alpha),
    \label{hbar}
\end{equation}
where
\begin{equation}
\beta_1 = \beta_+ + \sqrt{3} \beta_-, \quad  \beta_2 = \beta_+ - \sqrt{3} \beta_-, \quad \beta_3 = -2\beta_+.
\label{betai}
\end{equation}
Note that the spatial metric in equation (\ref{B3m}) describes the general Bianchi IX model. Two special cases are commonly distinguished in the literature \cite{book:R}, \cite{book:RS}. The diagonal (or non-rotating) case, where \( h_{ij} = \Bar{h}_{ij} \), is also known as the Mixmaster universe. In the symmetric (non-tumbling) case, rotation is restricted to a single axis, e.g., the \( z \)-axis with \( R \equiv R_z(\phi) \), resulting in a metric \( h_{ij} \) with a single off-diagonal component. Additionally, the closed FLRW model corresponds to a special case of the diagonal Bianchi IX universe, obtained by setting \( \beta_+ = \beta_- = 0 \). The axisymmetric Bianchi IX model is another special case, in which the space exhibits axial symmetry in addition to the \(\mathrm{SO}(3)\) invariance. This case is recovered by setting \( \beta_- = 0 \) in (\ref{hbar}), reducing the set of dynamical variables to \( \alpha \) and \( \beta_+ \).

The invariant basis \(\{e_i\}\) satisfies the condition $\mathcal{L}_{\xi_i} e_j = [\xi_i, e_j] = 0,$ where \(\xi_i\) are the Killing vectors generating the isometries and obey the commutation relations $[ \xi_i, \xi_j ] = C^k_{ij} \xi_k$. For a left-homogeneous space, the basis \(\{e_i\}\) is right-invariant and satisfies
\begin{equation}
   [ e_i, e_j ] = - C^k_{i j} e_k. 
   \label{grinv}
\end{equation}
The dual one-forms $\sigma^i$ satisfy the relations
\begin{equation}
    d \sigma^k = \frac12 C^k_{ij} \, \sigma^i \wedge \sigma^j. 
\end{equation}
The invariant basis vectors expressed explicitly in terms of the Euler angles is given by
\begin{equation}
\begin{aligned}
  &e_1 = \sin{\psi} \frac{\partial}{\partial \theta} + \cos{\psi} \left( \cot{\theta} \frac{\partial}{\partial \psi}  - \frac{1}{\sin{\theta}} \frac{\partial}{\partial \phi} \right),  \\
  &e_2 = \cos{\psi} \frac{\partial}{\partial \theta} - \sin{\psi} \left( \cot{\theta} \frac{\partial}{\partial \psi}  - \frac{1}{\sin{\theta}} \frac{\partial}{\partial \phi} \right),  \\
  &e_3 = \frac{\partial}{\partial \psi}.
\end{aligned}   
\label{eright}
\end{equation}
Accordingly, the one-forms $\sigma^i$ are given by
\begin{equation}
 \begin{aligned}
    &\sigma^1  = - \cos{\psi} \sin{\theta} d \phi + \sin{\psi} d \theta, \\
    &\sigma^2  = \sin{\psi}  \sin{\theta} d \phi + \cos{\psi} d \theta, \\
    &\sigma^3  =  \cos{\theta} d \phi + d \psi,
    \end{aligned}
\end{equation}
with the wedge product of
\begin{equation}
\sigma^1  \wedge \sigma^2  \wedge \sigma^3  =  \sin{\theta}  d \theta \wedge d \phi \wedge d \psi. 
\label{wedge}
\end{equation}
The left-invariant basis, $\tilde{e}_i \equiv \xi_i$, is related to the right-invariant basis by the matrix adjoint representation of the Lie group \cite{paper:Jan}. For Bianchi IX, the adjoint group is the group $\mathrm{SO}(3)$ itself. Consequently, the left- and right-invariant bases are related by rotation matrices $R\in \mathrm{SO}(3)$ as follows:
\begin{equation}
    \tilde{e}_j = {R^i}_j  {e}_i.
    \label{lr}
\end{equation}
Following the general procedure outlined in \cite{paper:UJ}, we introduce an orthonormal frame. This step is crucial for the vierbein formalism, which underpins the treatment of spinor fields in curved spacetime \cite{book:Birrell}, \cite{book:Parker}.

The spacetime metric $g_{\mu \nu}$ can be expressed in terms of the vierbein as
\begin{equation}
    g_{\mu \nu} (x) = \eta_{\hat{\alpha} \hat{\beta}} {\hat{h}}^{\hat{\alpha}}_{\ \mu}(x)  {\hat{h}}^{\hat{\beta}}_{\ \nu}(x),
\end{equation}
where $\eta_{\hat{\alpha} \hat{\beta}}$ is the Minkowski metric with signature $(-,+,+,+)$. Here, the Greek indices $\mu$, $\nu$ denote spacetime indices, while indices with hats, $\hat{\alpha}$, $\hat{\beta}$ refer to the local inertial frame. The vierbein (or tetrad) denoted by ${\hat{h}}^{\hat{\alpha}}_{\ \mu}(x)$ satisfies the relations 
\begin{equation}
 {{\hat{h}}_{\hat{\alpha}}}^{\ \mu} {\hat{h}}_{\hat{\beta}\mu} = \eta_{\hat{\alpha} \hat{\beta}} ,\quad   {{\hat{h}}^{\hat{\alpha}}}_{\ \mu} {{\hat{h}}_{\hat{\beta}}}^{\ \mu} = {\delta^{\hat{\alpha}}}_{\hat{\beta}}, \quad   {{\hat{h}}^{\hat{\alpha}}}_{\ \mu}  {\hat{h}}_{\hat{\alpha} \nu} = g_{\mu \nu},
\end{equation}
along with the duality relations:
\begin{equation}
   {{\hat{h}}_{\hat{\alpha}}}^{\ \mu}{{\hat{h}}^{\hat{\alpha}}}_{\ \nu} = {\delta^\mu}_\nu,\quad 
    {{\hat{h}}_{\hat{\alpha}}}^{\ \mu}{{\hat{h}}^{\hat{\beta}}}_{\ \mu} = {\delta^{\hat{\alpha}}}_{\hat{\beta}},
   \label{rect}
\end{equation}
where ${{\hat{h}}_{\hat{\alpha}}}^{\ \mu}$ is the reciprocal tetrad (or inverse vierbein). 

Introducing the one-forms \(\chi^{\hat{\alpha}}(x)\) of the local orthonormal frame via
\begin{equation}
  \chi^{\hat{\alpha}}(x) = {{\hat{h}}^{\hat{\alpha}}}_{\ \mu}(x) \, d x^\mu,
  \label{chialpha}
\end{equation}
the spatial metric \(h_{ij}\) can be expressed in terms of the vierbein components as
\begin{equation}
   h_{i j} (x) = \delta_{\hat{i} \hat{j}} \, h^{\hat{i}}_{\ i}(x) \, h^{\hat{j}}_{\ j}(x).
\end{equation}
Here, we defined
\[
    h^{\hat{i}}_{\ i} = {{\hat{h}}^{\hat{i}}}_{\ j} \, {\sigma^j}_i,
\]
where \({\sigma^j}_i\) relates the invariant one-forms \(\sigma^j\) to the coordinate basis via \( \sigma^j = {\sigma^j}_i \, dx^i \).

The one-forms of the orthonormal frame, given in (\ref{chialpha}) can then be expressed as
\begin{equation}
  \chi^{\hat{0}} = N d t,  \quad  \chi^{\hat{i}} = h^{\hat{i}}_{\ i} (N^i dt +  \sigma^i),
\end{equation}
where for Bianchi IX metric (\ref{B3m})
\begin{equation}
h^{\hat{i}}_{\ i} =  b^{k}  {R^{\hat{i}}}_i,  \quad h^{\hat{0}}_{\ 0} = N, \quad h^{\hat{0}}_{\ i} = 0,  \quad h^{\hat{i}}_{\ 0}  = b^{k}  {R^{\hat{i}}}_i N^i,
\label{vierh}
\end{equation}
where we introduced $b^k = e^{\tilde{\beta}_k/2}$ with $k =\hat{i}$ (the differing index notation is used deliberately to avoid implying a summation). 
Finally, the orthonormal basis vectors $\bar{e}_{\hat{i}}$ can be related to the invariant basis vectors $e_i$ through
\begin{equation}
  \bar{e}_{\hat{0}} = \frac{1}{N} (\partial_t - N^i e_i),  \quad  \bar{e}_{\hat{i}} = h_{\hat{i}}^{\ i} \, e_i.
  \label{orthe}
\end{equation}
Having established the Bianchi IX line element and introduced both invariant and orthonormal basis vectors, we proceed to the Lagrangian formulation of the Dirac field in this anisotropic spacetime.

\subsection{Lagrangian formulation}

We develop the Lagrangian formulation for the spinor field in Bianchi IX geometry. Previously, the dynamics of the Bianchi IX model coupled to a homogeneous spinor field were studied in \cite{paper:ORyan}, and the analysis was later extended to the Wheeler–DeWitt equation in this context in \cite{paper:Damour}. In contrast, since our focus is on solving and quantizing the spinor field, we do not impose the homogeneity restriction.

The Lagrangian density of the Dirac field in curved spacetime has the form
\begin{equation}
    \mathcal{L}_D = \sqrt{-g} \Bar{\Psi}\left( i \gamma^\mu\nabla_\mu  - m\right) \Psi,
\end{equation}
where $\Bar{\Psi} = \Psi^\dag \gamma^{\hat{0}}$. The generalized gamma matrices $\gamma^\mu$ are defined by 
\begin{equation}
 \gamma^\mu  = {{\hat{h}}_{\hat{\alpha}}}^{\ \mu} \gamma^{\hat{\alpha}},
 \label{gammamu}
\end{equation}
where $\gamma^{\hat{\alpha}}$ are the gamma matrices in Minkowski spacetime. Hence, the following anticommutation relations are satisfied
\begin{equation}
\left\{\gamma^\mu, \gamma^\nu\right\} = - 2 g^{\mu \nu}, \quad  \textnormal{and}  \quad \left\{\gamma^{\hat{\alpha}}, \gamma^{\hat{\beta}} \right \} = - 2 \eta^{\hat{\alpha} \hat{\beta}}.
\end{equation}
In the Weyl (or chiral) representation $\gamma^{\hat{\alpha}}$ matrices take the form 
\begin{equation}
  \gamma^{\hat{\alpha}} =  \left(\begin{matrix}
        0 & \sigma^{\hat{\alpha}} \\
        \Bar{\sigma}^{\hat{\alpha}} & 0
    \end{matrix} \right),
\label{gammabar}
\end{equation}
where we introduced the notation 
\begin{equation}
    \sigma^{\hat{\alpha}} = (I, \sigma^{\hat{i}}) , \quad \Bar{\sigma}^{\hat{\alpha}} = (I, - \sigma^{\hat{i}}), \quad {\hat{\alpha}} = \hat{0}, \hat{1}, \hat{2}, \hat{3}, 
    \label{sigmabar}
\end{equation}
with the Pauli matrices given by 
\begin{align}
\sigma^{\hat{1}}=\left(
\begin{matrix}
   0 & 1 \\
   1 & 0
\end{matrix}
\right),\  \
\sigma^{\hat{2}} =\left(
\begin{matrix}
   0 & - i \\
  i & 0 
\end{matrix}
\right),\  \
\sigma^{\hat{3}} =\left(
\begin{matrix}
  1 & 0\\
  0 & -1
\end{matrix}
\right).
\label{Pauli1}
\end{align}
The covariant derivative acting on the Dirac spinor is defined by
\begin{equation}
    \nabla_\mu \Psi =  (\partial_\mu + \Gamma_\mu) \Psi,
\end{equation}
where the connection 
\begin{equation}
    \Gamma_\mu = \frac12   \kappa_{\hat{\alpha} \hat{\beta} \mu} \Sigma^{\hat{\alpha} \hat{\beta}}, \quad \textnormal{with} \quad    \kappa_{\hat{\alpha} \hat{\beta} \mu}:=   {{\hat{h}}_{\hat{\alpha}\nu}} \left( \nabla_\mu {{\hat{h}}_{\hat{\beta}}}^{\ \nu}\right).
\label{CovDer}
\end{equation}
Here, $\Sigma^{\hat{\alpha} \hat{\beta}}$ are the generators of Lorentz group, i.e.
\begin{equation}
  \Sigma^{\hat{\alpha} \hat{\beta}} = \frac14 \left[\gamma^{\hat{\alpha}},\gamma^{\hat{\beta}}\right].
\label{Lorentz}
\end{equation}
Hence, the Lagrangian density can be written explicitly as follows:
\begin{equation}
\begin{aligned}
  \mathcal{L}_D =  \sqrt{-g} \Bar{\Psi}\left( i{{\hat{h}}_{\hat{\rho}}}^{\ \mu} \gamma^{\hat{\rho}}\partial_\mu   + \frac{i}{2} \kappa_{\hat{\alpha} \hat{\beta}\hat{\rho}}\gamma^{\hat{\rho}}  \Sigma^{\hat{\alpha} \hat{\beta}} - m \right)\Psi,
  \label{LagD}
  \end{aligned}
\end{equation}
where we introduced $   \kappa_{\hat{\alpha} \hat{\beta}\hat{\rho}} = {{\hat{h}}_{\hat{\rho}}}^{\ \mu} \kappa_{\hat{\alpha} \hat{\beta} \mu}$.
This quantity can be expressed in terms of commutation coefficients of the orthonormal basis, ${\gamma^{\hat{\sigma}}}_{\hat{\alpha} \hat{\beta}}$, defined by $ [\bar{e}_{\hat{\alpha}},\bar{e}_{\hat{\beta}}] = {\gamma^{\hat{\sigma}}}_{\hat{\alpha} \hat{\beta}}\, \bar{e}_{\hat{\sigma}}$.
These coefficients enter the expression
\begin{equation}
  \kappa_{\hat{\alpha} \hat{\beta}\hat{\rho}} 
  = \frac12 (\gamma_{\hat{\rho} \hat{\alpha} \hat{\beta}} + \gamma_{\hat{\beta} \hat{\alpha} \hat{\rho}} - \gamma_{\hat{\alpha} \hat{\beta}\hat{\rho}}),
  \label{kappa}
\end{equation}
where $\gamma_{\hat{\alpha} \hat{\beta} \hat{\rho}} = \eta_{\hat{\alpha} \hat{\tau}}{{\gamma}^{\hat{\tau}}}_{\hat{\beta} \hat{\rho}}$. 

The commutation coefficients can also be obtained from the equation for the dual one-forms $\chi^{\hat{\alpha}}$:
\begin{equation}
    d \chi^{\hat{\alpha}} = - \frac{1}{2} {\gamma^{\hat{\alpha}}}_{\hat{\beta} \hat{\gamma}}  \chi^{\hat{\beta}} \wedge \chi^{\hat{\gamma}}.
    \label{ctau}
\end{equation}
We begin by computing the relevant wedge products. When both indices are temporal, i.e., \(\hat{\beta} = \hat{\gamma} = \hat{0}\), the wedge product vanishes trivially:
\begin{equation}
    \chi^{\hat{0}} \wedge \chi^{\hat{0}} = 0.
\end{equation}
For mixed indices with \(\hat{\beta} = \hat{0}\) and \(\hat{\gamma} = \hat{l}\), where \(\hat{l} = \hat{1}, \hat{2}, \hat{3}\), the wedge product yields
\begin{equation}
    \chi^{\hat{0}} \wedge \chi^{\hat{l}} = N b^k {R^{\hat{l}}}_l\, dt \wedge \sigma^l, \quad \text{with } k = \hat{l}.
\end{equation}
For purely spatial indices, \(\hat{\beta} = \hat{j}\) and \(\hat{\gamma} = \hat{l}\), where \(\hat{j}, \hat{l} = \hat{1}, \hat{2}, \hat{3}\), the wedge product becomes:
\begin{equation}
\begin{aligned}
  \chi^{\hat{j}} \wedge \chi^{\hat{l}} 
     =\ & b^{m} b^{k} \left({R^{\hat{j}}}_j N^j    {R^{\hat{l}}}_l  - {R^{\hat{j}}}_l   {R^{\hat{l}}}_j N^j  \right) d t \wedge \sigma^l \\
    &+ b^{m} b^{k} {R^{\hat{j}}}_j   {R^{\hat{l}}}_l\,  \sigma^j \wedge  \sigma^l,
\end{aligned}
\label{chiw}
\end{equation}
where we identify \(\hat{j} = m\) and \(\hat{l} = k\). 

Next, we compute the exterior derivative \(d \chi^{\hat{\alpha}}\). For the temporal component \(\hat{\alpha} = \hat{0}\), we find
\begin{equation}
    d \chi^{\hat{0}}  = \dot{N}(t)\, dt \wedge dt = 0.
\end{equation}
For spatial components \(\hat{\alpha} = \hat{i}\), with \(\hat{i} = \hat{1}, \hat{2}, \hat{3}\), it reads
\begin{equation}
\begin{aligned}
d \chi^{\hat{i}} =\ & - {b}^{n}  {R^{\hat{i}}}_r {\left(\mathcal{J}_a\right)^r}_i  N^i(t)\, dt  \wedge \sigma^a \\
&+ \left(\dot{b}^n {R^{\hat{i}}}_i + b^{n}  {\dot{R}^{\hat{i}}}_i \right) dt \wedge \sigma^i \\
&+ {b}^{n}  {R^{\hat{i}}}_q {\left(\mathcal{J}_b\right)^q}_i \sigma^b \wedge \sigma^i,
\end{aligned}
\label{dchi}
\end{equation}
where \(n = \hat{i}\) and \(a, b = 1, 2, 3\). In deriving this, the following relation was used \cite{paper:Jan}:
\begin{equation}
    d {R^{\hat{i}}}_i = {R^{\hat{i}}}_r {\left(\mathcal{J}_a\right)^r}_i \sigma^a,
\end{equation}
where \(\mathcal{J}_a\) are the generators of \(\mathrm{SO}(3)\), satisfying the commutation relations \([ \mathcal{J}_i, \mathcal{J}_j ] = C^k_{ij} \mathcal{J}_k\).

Substituting the above relations into (\ref{ctau}) yields
\begin{equation}
\begin{aligned}
&{{\gamma}^{\hat{0} }}_{\hat{\beta} \hat{\gamma}} = 0 \quad \textnormal{for any} \ \hat{\beta}, \hat{\gamma},\\
&  {{\gamma}^{\hat{i} }}_{\hat{l} \hat{j}} 
  = {b}^{n} (b^{m})^{-1} (b^{k})^{-1} \varepsilon_{ \hat{i} \hat{j} \hat{l}},  \\
&  {{\gamma}^{\hat{i} }}_{\hat{0} \hat{l}}  = \frac{ (b^{k})^{-1}}{N} \Bigg[ 
    {b}^{n}  {R^{\hat{i}}}_q {\left(\mathcal{J}_j\right)^q}_l  N^j  - \left(\dot{b}^n {R^{\hat{i}}}_l + b^{n}  {\dot{R}^{\hat{i}}}_l \right) \Bigg] {R^l}_{\hat{l}},
\end{aligned}
\end{equation}
where $n = \hat{i}, \ m = \hat{j}, \ k = \hat{l}$.

Taking the gauge $N^i = 0$ for convenience and using the above relations, an explicit calculation of $\kappa_{\hat{\alpha} \hat{\beta}\hat{\rho}}$ from (\ref{kappa}) yields the following non-vanishing contributions to the Lagrangian density:
\begin{equation}
\begin{aligned}
\frac{i}{2} \kappa_{\hat{\alpha} \hat{\beta}\hat{\rho}}\gamma^{\hat{\rho}}  \Sigma^{\hat{\alpha} \hat{\beta}} 
= \ & \frac{i}{2}  \left( \kappa_{\hat{0} \hat{1}\hat{1}} +  \kappa_{\hat{0} \hat{2}\hat{2}} + \kappa_{\hat{0} \hat{3}\hat{3}} \right)\gamma^{\hat{0}} \\
&+ i \left( \kappa_{\hat{1} \hat{2}\hat{3}} -  \kappa_{\hat{1} \hat{3}\hat{2}}  +  \kappa_{\hat{2} \hat{3}\hat{1}}\right) \gamma^{\hat{1}}  \Sigma^{\hat{2} \hat{3}}\\
&+ i  \left(\kappa_{\hat{1} \hat{2}\hat{0}}\gamma^{\hat{0}}  \Sigma^{\hat{1} \hat{2}} + \kappa_{\hat{1} \hat{3}\hat{0}}\gamma^{\hat{0}}  \Sigma^{\hat{1} \hat{3}} + \kappa_{\hat{2} \hat{3}\hat{0}}\gamma^{\hat{0}}  \Sigma^{\hat{2} \hat{3}} \right).
\label{Lgm}
\end{aligned}
\end{equation}
Finally, the Lagrangian density for the Dirac spinor field in Bianchi IX takes the form
\begin{equation}
\begin{aligned}
  \mathcal{L}_D = \ & N e^{3 \alpha} \Bar{\Psi}\Bigg\{ i \left(\gamma^\mu  \partial_\mu- \frac{3}{2N} \dot{\alpha} \gamma^{\hat{0}} \right)  - m 
\\
&+ i F(\alpha, \beta_+,\beta_-) \gamma^{\hat{1}}  \Sigma^{\hat{2} \hat{3}}
  -  \frac{i}{N}\Bigg[  \left(1 +\frac{I_{\hat{3}}}{4}\right)^{\frac12}  \Omega_{\hat{3}}   \gamma^{\hat{0}}  \Sigma^{\hat{1} \hat{2}} \\
 & +   \left(1 +\frac{I_{\hat{2}}}{4}\right)^{\frac12} \Omega_{\hat{2}}  \gamma^{\hat{0}}  \Sigma^{\hat{3} \hat{1}} 
  +   \left(1 +\frac{I_{\hat{1}}}{4}\right)^{\frac12}  \Omega_{\hat{1}}  \gamma^{\hat{0}}  \Sigma^{\hat{2} \hat{3}} \Bigg] \Bigg\}\Psi,
  \label{SLag1} 
\end{aligned} 
\end{equation}    
where we introduced
\begin{equation}
\resizebox{1.\hsize}{!}{$
F(\alpha, \beta_+, \beta_-) := \frac12e^{-\alpha} \left[e^{- 4 \beta_+} + e^{2 \beta_+ + 2 \sqrt{3} \beta_-}  + e^{2 \beta_+ - 2 \sqrt{3} \beta_-}\right],
$}
\label{FB}
\end{equation}
and the components of the ``angular velocity'', defined as
\begin{equation}
    \Omega_{\hat{1}} = {\omega^{\hat{2}}}_{\hat{3}}, \ \Omega_{\hat{2}} = {\omega^{\hat{3}}}_{\hat{1}}, \ \Omega_{\hat{3}} = {\omega^{\hat{1}}}_{\hat{2}}
\end{equation}
with 
\begin{equation}
    {\omega^{\hat{k}}}_{\hat{n}} = {{\dot{R}}^{\hat{k}}}_{\,\,\,\hat{i}} {R^{\hat{i}}}_{\hat{n}}.
\end{equation}
Explicitly, they take the form
\begin{equation}
\begin{aligned}
    & {\omega^{\hat{2}}}_{\hat{3}} = \sin{\psi} \dot{\theta} - \sin{\theta} \cos{\psi} \dot{\phi}, \\
    &{\omega^{\hat{3}}}_{\hat{1}} = \sin{\psi} \sin{\theta}  \dot{\phi} + \cos{\psi} \dot{\theta}, \\
    & {\omega^{\hat{1}}}_{\hat{2}} =  \dot{\psi} +  \cos{\theta} \dot{\phi}.
    \label{omega}
\end{aligned}
\end{equation}
The quantities $I_{\hat{i}}$, referred to as ``moments of inertia'', are given by 
\begin{equation}
\begin{aligned}
     & I_{\hat{1}} = 4 \sinh^2{\left(3\beta_+ - \sqrt{3}\beta_- \right)} ,  \\
     & I_{\hat{2}} = 4  \sinh^2{\left(3 \beta_+ + \sqrt{3}\beta_-\right)},  \\
     & I_{\hat{3}} = 4 \sinh^2{\big(2\sqrt{3} \beta_-\big)}.
    \label{I1I2I3} 
\end{aligned}
\end{equation}
The introduction of the ``moments of inertia'' is motivated by the fact that the rotational contribution in the gravitational Lagrangian density is analogous to the kinetic energy of an asymmetric top (see, e.g., \cite{paper:Damour}, \cite{paper:Nick}).

In the Lagrangian density \eqref{SLag1}, the first two terms in brackets are, as expected, also present in isotropic FLRW models \cite{paper:ParkerS}. However, in the Bianchi IX background, additional coupling terms appear: a geometry–spinor coupling potential, 
\( \propto F(\alpha, \beta_+, \beta_-) \gamma^{\hat{1}}  \Sigma^{\hat{2} \hat{3}} \), arising from its anisotropy; a spinor–``angular velocity'' coupling term, 
\( \propto \Omega_{\hat{i}}  \gamma^{\hat{0}}  \Sigma^{\hat{j} \hat{k}} \),  resulting from the global rotation of the model. Later, the contributions of these terms to the particle–antiparticle spectral asymmetry will be explored by analyzing the solutions of the Dirac equation in this background in Sec.~\ref{sol}.

\subsection{Dirac field equation} \label{eomWD}

The Dirac equation in general Bianchi IX model derived from the Euler-Lagrange equation using the spinor Lagrangian (\ref{SLag1}) takes the form
\begin{widetext}
\begin{equation}
\begin{aligned}
 \Bigg\{ &i \left( \frac{1}{N}\gamma^{\hat{0}} \partial_0 +{h_{\hat{i}}}^i\gamma^{\hat{i}} e_i - \frac{3}{2N} \dot{\alpha} \gamma^{\hat{0}} \right)   + i F(\alpha, \beta_+, \beta_-)\gamma^{\hat{1}}  \Sigma^{\hat{2} \hat{3}} - m  \\
  &-  \frac{i}{N}\Bigg[ \left(1 +\frac{I_{\hat{3}}}{4}\right)^{\frac12}  \Omega_{\hat{3}} \,\gamma^{\hat{0}}  \Sigma^{\hat{1} \hat{2}}  +  \left(1 +\frac{I_{\hat{2}}}{4}\right)^{\frac12} \Omega_{\hat{2}} \,\gamma^{\hat{0}}  \Sigma^{\hat{3} \hat{1}} 
  +   \left(1 +\frac{I_{\hat{1}}}{4}\right)^{\frac12}  \Omega_{\hat{1}}\, \gamma^{\hat{0}}  \Sigma^{\hat{2} \hat{3}} \Bigg] \Bigg\}\Psi = 0.
  \label{geom}
  \end{aligned}
\end{equation}
\end{widetext}
The special Bianchi IX cases discussed earlier can be recovered from the general Dirac equation \eqref{geom}. For the diagonal case, the rotational coupling terms in the last bracket vanish entirely. In the symmetric case, only the term involving 
$\Omega_{\hat{3}}$ contributes. The non-rotating and rotating axisymmetric cases arise from the diagonal and symmetric cases, respectively, by further imposing $\beta_- =0$.

The Dirac spinor $\Psi$ can be written in the form
\begin{equation}
    \Psi =  \left( 
    \begin{matrix}
        \Psi_L \\
        \Psi_R
    \end{matrix}
    \right),
    \label{psiLR}
\end{equation}
where $\Psi_L$ and $\Psi_R$ are the left- and right-handed Weyl spinors, respectively.
Furthermore, using the gamma matrices in the Weyl representation \eqref{gammabar}, the Dirac equation can be expressed in terms of Pauli matrices as
\begin{equation}
 \left(
\begin{matrix}
  - m & D_R
  \\
D_L
 & - m 
\end{matrix}
\right) \left( 
    \begin{matrix}
        \Psi_L \\
        \Psi_R
    \end{matrix}
    \right)  =0,
\label{Weom}
\end{equation}
where 
\begin{equation}
\resizebox{1.\hsize}{!}{$
 \begin{aligned}
    &D_R = i {h_{\hat{\alpha}}}^\mu \sigma^{\hat{\alpha}} e_\mu- \frac{3 i}{2N} \dot{\alpha} + \frac12 F(\alpha, \beta_{+},\beta_{-})- \frac{1}{2N}  \sum_{\hat{l}}\bar{I}_{\hat{l}} \Omega_{\hat{l}} \sigma^{\hat{l}}, \\
   &D_L =  i {h_{\hat{\alpha}}}^\mu  \Bar{\sigma}^{\hat{\alpha}} e_\mu - \frac{3 i}{2N} \dot{\alpha}  - \frac12F(\alpha, \beta_{+},\beta_{-}) - \frac{1}{2N}  \sum_{\hat{l}}\bar{I}_{\hat{l}}  \Omega_{\hat{l}} \sigma^{\hat{l}}.
\end{aligned}$}
\end{equation}
The operators $D_R$ and $D_L$ act on $\Psi_R$ and $\Psi_L$ spinors, respectively. Here, for convenience, we have introduced $\bar{I}_{\hat{i}}: = \left(1 + I_{\hat{i}}/4\right)^{1/2}$. 

To solve Eq.~\eqref{Weom}, we proceed by performing a harmonic analysis of the spinor field on the Bianchi IX geometry.

\section{Harmonic analysis of spinor fields in Bianchi IX geometry}  \label{HA}

Harmonic analysis of spinor fields in a fixed Bianchi IX background can be formulated using the group-theoretic structure of its spatial symmetry group, $\mathrm{SO(3)}$. Appendix~\ref{irr} reviews the relevant unitary irreducible representations of the double cover $\mathrm{SU(2)}$ of $\mathrm{SO(3)}$—namely, the spinor representation, the representation on functions, and their tensor product, which underlies the mode decomposition.

In their work on the fixed Mixmaster universe, Dowker and Pettengill~\cite{paper:Dowker} formulated a harmonic analysis for spinor fields, building on earlier scalar-field results by Hu~\cite{paper:HuI}. They demonstrated that the second-order Dirac equation can be recast as the eigenvalue problem for the Hamiltonian of an \textit{ideal asymmetric top}—an asymmetric top with intrinsic spin. Hence, in this framework, constructing a mode decomposition reduces to determining the eigenstates of the ideal-top Hamiltonian.

In the present work, we extend this analogy beyond the Mixmaster (diagonal) model to encompass the general Bianchi IX geometry. In addition, we complete the discussion by highlighting two natural choices of commuting operator sets and their corresponding eigenbases, based on group-theoretic considerations.

\subsection{Ideal top analogy}

The second-order equation for the Dirac spinor field $\Psi$ is given by
\begin{equation}
\left[ \nabla_\mu \nabla^\mu + R/4 - M^2 \right] \Psi = 0,
\label{sec}
\end{equation}
where
\begin{equation}
\frac{R}{4} = \frac12 R_{\mu \nu \lambda \sigma } \Sigma^{\mu\nu} \Sigma^{\lambda\sigma}, \quad \textnormal{with} \quad \Sigma^{\mu\nu} := \frac14 [\gamma^\mu, \gamma^\nu].
\end{equation}
Here, $R_{\mu \nu \lambda \sigma }$	denotes the Riemann curvature tensor of the Bianchi IX spacetime, and $R$ is the corresponding Ricci scalar.
Expressing $\Psi$ in terms of left- and right-handed Weyl spinors, the second-order equation for each Weyl spinor becomes \cite{paper:Dowker}
\begin{equation}
\left[ \nabla_\mu \nabla^\mu + \frac12 R_{ \mu \nu \lambda \sigma} \Sigma_{L,R}^{\mu\nu} \Sigma_{L,R}^{\lambda\sigma} - M^2 \right] \Psi_{L,R} = 0,
\label{secLR}
\end{equation}
where $\Sigma_{L,R}^{\mu\nu}$ can be written as
\begin{equation}
\begin{aligned}
  \Sigma^{\mu\nu}_{L,R} = {\hat{h}_{\hat{\alpha}}}^{\ \mu} {\hat{h}_{\hat{\beta}}}^{\ \nu} \Sigma^{\hat{\alpha}\hat{\beta}}_{L,R},
\end{aligned}
\end{equation}
with $\Sigma^{\hat{\alpha}\hat{\beta}}_{L,R}$ denoting the Lorentz group generators in the $\left(\frac12,0\right)$ and $\left(0, \frac12\right)$ representations, respectively \cite{book:Srednicki}.

For a fixed Bianchi IX background, where the metric parameters $\alpha$, $\beta_+$, $\beta_-$, and $\Omega_i$ are treated as constants, the generators $\Sigma^{\hat{\alpha}\hat{\beta}}_{L,R}$ are proportional to the components of the spin angular momentum operator, $j_i = \sigma_i/2$. Hence, the curvature term can be expressed as
\begin{equation}
\frac12 R_{ \mu \nu \lambda \sigma} \Sigma_{L,R}^{\mu\nu} \Sigma_{L,R}^{\lambda\sigma}  = \sum_i c_i {\tilde{j}}^2_i,
\label{RSS}
\end{equation}
where the $c_i$ are constants depending on the Bianchi IX metric parameters that can be determined through explicit calculations. The operators ${\tilde{j}}_{i}$ represent the components of spin angular momentum in the space-fixed frame and are related to the body-fixed components via ${\tilde{j}}_{i} = {R^m}_i \, j_m$. In the Mixmaster case, the relation (\ref{RSS}) is instead proportional to the body-fixed spin angular momentum \cite{paper:Dowker}.

Furthermore, the term $\nabla_\mu \nabla^\mu$ in Eq.~(\ref{secLR}) can be written in the $N^i = 0$ gauge using the spatial three-metric~(\ref{B3m}) as:
\begin{equation}
\begin{aligned}
    \nabla_\mu  \nabla^\mu 
  =\ &   -\frac{1}{N^2(t)} \partial_t^2  + {R^a}_m  {R^b}_n \Bar{h}^{m n}\\
 & \times (e_a e_b + e_a \,{\sigma_b}^j \Gamma_j  + {\sigma_a}^i \Gamma_i \, e_b + {\sigma_a}^i {\sigma_b}^j  \Gamma_i \Gamma_j).
\label{nn}
\end{aligned}
\end{equation}
Using the relation~\eqref{lr} between the left- and right-invariant bases, along with Eq.~(\ref{JRL}), one finds
\begin{equation}
     {R^a}_m {R^b}_n  {\Bar{h}}^{m n} e_a e_b 
     = \sum_i \tilde{\rho}_i \tilde{L}_i^2,
\end{equation}
where are $\tilde{\rho}_i$ constants. 

Moreover, the covariant derivative acting on $\Psi_{L,R}$ can be introduced using Eq.~\eqref{CovDer}:
\begin{equation}
\begin{aligned}
  \Gamma^{L,R}_i &= \frac12 \kappa_{\hat{\alpha} \hat{\beta} i} \Sigma_{L,R}^{\hat{\alpha} \hat{\beta}}.
\end{aligned}
\end{equation}
The corresponding contributions to Eq.~(\ref{secLR}) can then be written as
\begin{equation}
{R^a}_m  {R^b}_n \Bar{h}^{m n} {\sigma_a}^i {\sigma_b}^j  \Gamma_i \Gamma_j  = \sum_i \tilde{\upsilon}_i \, {\tilde{j}}^2_{i},
\end{equation}
and
\begin{equation}
\begin{aligned}
  {R^a}_m  {R^b}_n \Bar{h}^{m n} ( e_a \,{\sigma_b}^j \Gamma_j  + {\sigma_a}^i \Gamma_i \, e_b ) = \sum_i  \xi_i  \tilde{L}_i \tilde{j}_i,
  \label{coupling}
\end{aligned}
\end{equation}
where $\tilde{\upsilon}_i$ and $\xi_i$ are constants, and Eq.~(\ref{coupling}) represents the spin-orbit coupling.
Introducing total angular momentum in the body-fixed and space-fixed frames as
\begin{equation}
J_i = L_i + j_i, \quad \tilde{J}_i = \tilde{L}_i + \tilde{j}_i,
\end{equation}
which are related via ${\tilde{J}}_i = {R^m}_i\, J_m$, the Hamiltonian eigenvalue equation follows from Eq.~(\ref{secLR}):
\begin{equation}
H \Psi_{L,R} = (E^2 - M^2)\Psi_{L,R},
\label{Hamilton}
\end{equation}
with the Hamiltonian
\begin{equation}
H = \sum_i \left( \tilde{\lambda}_i {\tilde{L}}^2_i + \tilde{\mu}_i {\tilde{j}}^2_i + \tilde{\nu}_i {\tilde{J}}^2_i \right),
\label{asitB}
\end{equation}
where $\tilde{\lambda}_i$, $\tilde{\mu}_i$, and $\tilde{\nu}_i$ are constants reflecting contributions from each angular momentum component. 

From the general Bianchi IX result obtained here, the Mixmaster case of~\cite{paper:Dowker} is recovered by replacing the space-fixed angular momentum components with their body-fixed counterparts, yielding the Hamiltonian of an ideal asymmetric top:
\begin{equation}
H = \sum_i \left( \lambda_i L^2_i + \mu_i j^2_i + \nu_i J^2_i \right).
\label{asit}
\end{equation}
As a special case, the ideal spherical top Hamiltonian simplifies to 
\begin{equation}
H = \lambda L^2 + \mu j^2 + \nu J^2,
\label{sit}
\end{equation}
where each set of angular momentum components shares a common coefficient.

\subsection{Eigenstates of the ideal symmetric top}\label{HAs}

We begin by considering the ideal spherical top Hamiltonian in (\ref{sit}). This Hamiltonian commutes with the operators $J^2$, $L^2 = \tilde{L}^2$, $j^2$, $J_3$, and $\tilde{L}_3$. The choice of a particular eigenbasis corresponds to selecting a set of operators that commute with the Hamiltonian; thus, the eigenstates coincide with those derived from the group-theoretic discussion in Appendix~\ref{irr}, as given in either Eq.~\eqref{A15} or Eq.~\eqref{A17}.  

For the commuting set $L^2 = {\tilde{L}}^2$, $j^2$, $L_3$, $j_3$, and $\tilde{L}_3$, the eigenstates take the form
\begin{equation}
   |l \, s;  m \,( n - n_s) \, n_s\rangle =|l, m , n - n_s \rangle|s\, n_s \rangle,
   \label{symln}
\end{equation}
where $|s\, n_s \rangle$ denote the spinor states defined in Eq.~\eqref{speig}, and $|l, m , n_l \rangle$, with $n_l = n - n_s$, are the eigenstates of the representation on functions, as introduced in Eq.~\eqref{orthket}.

Alternatively, for the commuting set $J^2$, $L^2 = {\tilde{L}}^2$, $j^2$, $J_3$, $\Tilde{L}_3$, the eigenstates are given by $| l \,s ; j \, m \, n\big>$ and can be expanded in terms of the previous basis as
\begin{equation}
   | l \,s ; j \, m \, n\big> = \sum_{ n_s = \pm \frac12} C^{j n}_{l, n - n_s; s n_s}  |l, m, n - n_s \big>  | s \, n_s \big>,
   \label{symjn}
\end{equation}
with $j =  l\pm \frac12$, and $C^{j n}_{l, n - n_s; s n_s}$ the corresponding Clebsch–Gordan coefficients. Explicit values of these coefficients are given in Table~\ref{table:species5} in Appendix~\ref{irr}.

In the space-fixed frame, the Hamiltonian of the ideal spherical top follows as a special case of the asymmetric top Hamiltonian in Eq.~\eqref{asitB}:
\begin{equation}
H = \tilde{\lambda} {\tilde{L}}^2 + \tilde{\mu} {\tilde{j}}^2 + \tilde{\nu} {\tilde{J}}^2.
\end{equation}
This Hamiltonian commutes with ${\tilde{J}}^2$, $L^2 = {\tilde{L}}^2$, ${\tilde{j}}^2$, $\tilde{J}_3$, and $L_3$.  The eigenbasis of the space-fixed Hamiltonian can be obtained directly from the body-fixed eigenbasis by swapping the quantum numbers $m$ and $n$, reflecting the change from body-fixed to space-fixed frame. In particular, the eigenstates can be written as
\begin{equation}
   |l \, s;  (m - m_s) \,n \big> 
   =|l, m - m_s ,n \big> \big|s\, m_s \big>, 
   \label{symlm}
\end{equation}
where $m_s = \pm \frac12$. Alternatively, one may choose the eigenbasis
\begin{equation}
   | l \,s ; j \, m \, n\big> = \sum_{ m_s = \pm \frac12} C^{j m}_{l, m - m_s; s m_s}  |l, m - m_s, n \big>  | s \, m_s \big>.
   \label{symjm}
\end{equation}

The ideal symmetric top corresponds to the axisymmetric Bianchi IX models in our framework. In this setting, the body-fixed Hamiltonian corresponds to the non-rotating case, while the space-fixed Hamiltonian corresponds to the rotating case. The eigenstates of the ideal symmetric top are identical to those of the ideal spherical top constructed above. As discussed in Appendix~\ref{rottop}, the standard symmetric top shares the same rotational eigenstates as the spherical top, and this correspondence naturally extends to the ideal symmetric top. In particular, when solving the Dirac equation for axisymmetric Bianchi IX models in the next section, we employ the eigenbasis of Eq.~\eqref{symln} for the non-rotating case and Eq.~\eqref{symlm} for the rotating case, as convenient choices for the mode expansion.

The eigenstates of the ideal asymmetric top, relevant for the general Bianchi IX geometry, are also considered. In Appendix~\ref{rottop}, we review the eigenstates of the standard asymmetric top, expressed as linear combinations of symmetric top eigenstates. In addition, a symmetry-adapted eigenbasis, exploiting the intrinsic symmetries of the top, is discussed. Following the same construction, these results are generalized to the ideal asymmetric top. Due to the mathematical complexity of this derivation and the focus of the present work on axisymmetric Bianchi IX models, the full discussion is presented in Appendix~\ref{aTop}.

\section{Solutions of the Dirac Equation in Axisymmetric Bianchi IX Cosmology}\label{sol}

\subsection{Non-rotating axisymmetric case}\label{nonax}

The axisymmetric Bianchi IX model is obtained by imposing axial symmetry on the diagonal Bianchi IX metric. This corresponds to setting $\beta_- = 0$, which yields $\beta_{\hat{1}} = \beta_{\hat{2}} = \beta_+$ and $\beta_{\hat{3}} = -2\beta_+$ in Eq.~\eqref{betai}. The resulting spatial line element takes the form
\begin{equation}
    d l^2  
    =   e^{2 (\beta_+ + \alpha)}  \left[\left( \sigma^1 \right)^2 +\left( \sigma^2 \right)^2 \right] +   e^{2 (- 2\beta_+ + \alpha)}  \left( \sigma^3 \right)^2.
\end{equation}
From the vierbein given in Eq.~\eqref{vierh}, the corresponding inverse vierbein, which enters into the Dirac equation Eq.~\eqref{Weom}, simplifies to
\begin{equation}
    {h_{\hat{i}}}^i = (b^{k})^{-1} {\delta^i}_{\hat{i}}, \quad {h_{\hat{0}}}^0  = \frac{1}{N}, \quad {h_{\hat{0}}}^i = 0, \quad {h_{\hat{i}}}^0 = 0, \quad k =\hat{i}.
    \label{axisymv}
\end{equation}
Specializing the Dirac equation to the axisymmetric Bianchi IX background requires dropping the rotational contributions $\propto \Omega_{\hat{l}}$ in Eq.~\eqref{Weom}. It is convenient to work in conformal time, achieved by setting $N = e^{\alpha}$. In the present analysis, we consider a fixed background, taking $\alpha$ and $\beta_+$ to be constants. The resulting set of coupled equations reads
\begin{equation}
\begin{aligned}
  \left[ i {\sigma}^{\hat{0}} \partial_\eta + i e^{- (\beta_{\hat{i}} +\alpha)}\sigma^{\hat{i}} e_i   + \frac12 \tilde{F}(\alpha, \beta_+)   \right] \Psi_R &= M \Psi_L,\\
  \left[  i {\sigma}^{\hat{0}}  \partial_\eta - i e^{- (\beta_{\hat{i}} +\alpha)}{\sigma}^{\hat{i}} e_i - \frac12 \tilde{F}(\alpha, \beta_+) \right] \Psi_L &= M \Psi_R,
  \label{75}
  \end{aligned} 
\end{equation}
where $\tilde{F}(\alpha, \beta_+)$ is obtained from $F(\alpha, \beta_+, \beta_-)$ defined in Eq.~\eqref{FB} by setting $\beta_- = 0$, i.e.,
\begin{equation}
     \tilde{F}(\alpha, \beta_{+}): = \frac12 e^{-\alpha} \left[e^{- 4 \beta_+} + 2 e^{2 \beta_+} \right].
     \label{Ftilde}
\end{equation}
In a fixed spacetime, the absence of time dependence in the metric allows separation of variables in the Dirac equation. Accordingly, we seek solutions of the form
\begin{equation}
    \Psi (\eta,\phi,\theta,\psi) =  \left( 
    \begin{matrix}
        \Psi_L \\
        \Psi_R
    \end{matrix}
    \right) = N(\eta)  \left( 
    \begin{matrix}
        w_L(\phi,\theta,\psi) \\
        w_R(\phi,\theta,\psi)
    \end{matrix}
    \right),
    \label{dst}
\end{equation}
where $N(\eta)$ captures the temporal dependence, and $w_{L,R}(\phi,\theta,\psi)$ are two-spinors that depend on the spatial coordinates. 

Substituting the decomposition \eqref{dst} into \eqref{75} and separating variables, the temporal factor satisfies
\begin{equation}
     i  \frac{\left(\partial_\eta  N(\eta)\right)}{N(\eta)}  =  E  \quad \Rightarrow \quad N(\eta) = e^{ - i E \eta},
     \label{nst}
\end{equation}
where $E$ is a separation constant. For the spatial part, making use of Eq.~\eqref{JRL} and introducing the ladder operators $L_\pm = L_{\hat{1}} \pm i L_{\hat{2}}$, and the Pauli matrices in~\eqref{Pauli1}, the equations take the form
\begin{widetext}
\begin{equation}
\begin{aligned}
    \left(
\begin{matrix}
E + \frac12 \tilde{F}(\alpha, \beta_{+}) +  e^{2\beta_+ - \alpha}\, L_{\hat{3}}   &  e^{- (\beta_+ +\alpha)} \, L_-\\
    e^{- (\beta_+ + \alpha)}\, L_+ &  E + \frac12 \tilde{F}(\alpha, \beta_{+}) -  e^{2\beta_+ -\alpha} \,L_{\hat{3}} 
\end{matrix}
\right) w_R = M  w_L,
\label{79}
  \end{aligned}  
\end{equation}
and 
\begin{equation}
\begin{aligned}
  \left(
\begin{matrix}
  E - \frac12 \tilde{F}(\alpha, \beta_{+}) -  e^{2\beta_+ -\alpha}\, L_{\hat{3}} & -  e^{- (\beta_+ + \alpha)}\, L_-\\
  -  e^{- (\beta_+ + \alpha)} \,L_+ &  E - \frac12 \tilde{F}(\alpha, \beta_{+}) +   e^{2\beta_+ -\alpha}\, L_{\hat{3}} 
\end{matrix}
\right) w_L = M w_R.
\label{80}
  \end{aligned} 
\end{equation}
\end{widetext}
Expanding each two-spinor in terms of the ideal symmetric top eigenbasis (\ref{symln}), we look for solutions in following form
\begin{equation}
  w^{nlm}_L =  \left( \begin{matrix}
    w_1 \\
    w_2
   \end{matrix} \right) = \left( \begin{matrix}
        f^1_{nlm} |l , m ,n - \frac12\big> \\ 
     f^2_{nlm}  |l , m ,n + \frac12\big>
    \end{matrix} \right),
    \label{81}
\end{equation}
and
\begin{equation}
  w^{nlm}_R =  \left( \begin{matrix}
    w_3 \\
    w_4
   \end{matrix} \right) = \left( \begin{matrix}
     f^3_{nlm} |l , m ,n - \frac12\big> \\ 
     f^4_{nlm}  |l , m ,n + \frac12\big> 
    \end{matrix} \right),
    \label{82}
\end{equation}
where $f^{i}_{nlm}$ ($i=1,2,3,4$) denote the expansion coefficients. Substituting these expansions into Eqs.~\eqref{79} and \eqref{80}, using the eigenvalue relations (\ref{Llmn}) and eliminating the expansion coefficients, yields the following matrix equation:
\begin{equation}
    A := \left(\begin{matrix}
    a_{11} & a_{12}\\
    a_{21} & a_{22}
    \end{matrix}\right) = 0,
\end{equation}
where
\begin{equation}
\begin{aligned}
  a_{11}  = \ & E^2 -  M^2 -  e^{-2(\alpha + \beta_+)}  \left[ n^2 e^{6 \beta_+} + \alpha^2_{n + \frac12}\right] \\
  &- \frac{1}{16}e^{-2(\alpha +4 \beta_+)} - \frac12  n \, e^{-2(\alpha + \beta_+)}, \\
  a_{12} =\ &  a_{21} = - \frac12 e^{- 5 \beta_+ - 2 \alpha} \, \alpha_{n + \frac12}, \\
  a_{22} = \ & E^2  -  M^2 -  e^{-2(\alpha + \beta_+)}  \left[ n^2 e^{6 \beta_+} + \alpha^2_{n + \frac12}\right]\\
  &- \frac{1}{16}e^{-2(\alpha +4 \beta_+)} + \frac12  n \,e^{-2(\alpha + \beta_+)}.
\end{aligned}
\end{equation}
Solving this equation, by employing the characteristic equation, yields the energy eigenvalues:
\begin{equation}
  E^s_{1} = - \sqrt{M^2 +\left[ G(\alpha,\beta_+) \pm \tilde{\omega}_{nl} (\alpha, \beta_+)\right]^2},
  \label{7134}
\end{equation}
and
\begin{equation}
  E^s_{2} =  \sqrt{M^2 +\left[G(\alpha,\beta_+) \pm \tilde{\omega}_{nl} (\alpha, \beta_+)\right]^2},
  \label{7135}
\end{equation}
with $s=1,2$. Here, we introduced functions
\begin{equation}
G(\alpha,\beta_+): 
 =  \frac14 e^{- 4\beta_+  - \alpha}, 
 \label{Gtilde}
\end{equation}
and 
\begin{equation}
  \tilde{\omega}_{n l} (\alpha, \beta_+) = e^{- (\beta_+ + \alpha)} \left[ n^2 e^{6\beta_+}   +   \alpha^2_{n + \frac12}  \right]^{1/2}.
  \label{omeganl}
\end{equation}
The negative eigenvalues correspond to particles and the positive eigenvalues correspond to antiparticles (this eigenvalue assignment follows from the choice of metric signature). The corresponding eigenspinors can be obtained in a straightforward manner, but we do not present them here, as our focus is on the energy eigenvalues.

We observe that for the Dirac spinors, the spectrum exhibits well-defined positive and negative energy eigenvalues, unlike the case for Weyl spinors \cite{paper:Gib1}, and thus no level crossing occurs (as expected due to charge conservation). Moreover, the energy eigenvalues of a given mode exhibit a spin-dependent splitting: depending on the spin index $s$, levels are either enhanced or suppressed for both particles and antiparticles; this splitting arises solely from the spatial anisotropy of the background.

\subsection{Rotating axisymmetric case}

The three-metric for the rotating axisymmetric Bianchi~IX model follows from the axisymmetric restriction $\beta_- = 0$ of Eq.~\eqref{B3m}, with rotation about a single axis:
\begin{equation}
    h\big|_{\beta_- = 0} = R_z^T(\phi)\,\Bar{h}\big|_{\beta_- = 0}\,R_z(\phi).  
\end{equation}
In the Dirac equation~\eqref{Weom}, the rotational contribution in the axisymmetric case reduces to the term proportional to the angular velocity component $\Omega_{\hat{3}}$, with $I_{\hat{3}}=0$ as follows from the definition of the moments of inertia~\eqref{I1I2I3}.

Applying the temporal and spatial decomposition (\ref{dst}) together with (\ref{nst}), the spatial part of the Dirac equation in the rotating axisymmetric background can be written in terms of the angular momentum operators. Using the relations between left- and right-invariant bases~\eqref{lr} and the definition of the space-fixed angular momentum operators~\eqref{JRL}, the resulting equations involve the space-fixed operators due to the presence of rotation. The spatial equations take the form:
\begin{widetext}
\begin{equation}
\left( 
\begin{matrix}
   E  +  \frac12 \tilde{F}(\alpha, \beta_{+})    -    e^{2\beta_+ - \alpha}   \tilde{L}_{\hat{3}} -  \frac{1}{2} e^{-\alpha} \Omega_{\hat{3}}  &  - e^{-(\beta_+ + \alpha)} \tilde{L}_- \\
    - e^{- (\beta_+ + \alpha)} \tilde{L}_+ &  E  +  \frac12 \tilde{F}(\alpha, \beta_{+}) +  e^{2\beta_+ - \alpha}  \tilde{L}_{\hat{3}}  +  \frac{1}{2} e^{-\alpha} \Omega_{\hat{3}}
\end{matrix}
\right)   w_R = M    w_L,
\label{111}
\end{equation}
and
\begin{equation}
  \left( 
\begin{matrix}
     E  - \frac12 \tilde{F}(\alpha, \beta_{+}) + e^{2\beta_+ - \alpha} \tilde{L}_{\hat{3}}  - \frac{1}{2} e^{-\alpha} \Omega_{\hat{3}} &       e^{-(\beta_+ + \alpha)}   \tilde{L}_- \\
    e^{- (\beta_+ + \alpha)}  \tilde{L}_+ &  E  - \frac12 \tilde{F}(\alpha, \beta_{+})    -   e^{2\beta_+ - \alpha}   \tilde{L}_{\hat{3}}  +  \frac{1}{2} e^{-\alpha} \Omega_{\hat{3}}
\end{matrix}
\right)    w_L = M   w_R.
\label{222}
\end{equation}
\end{widetext}
For the rotating case, the expansion of $w_L$ and $w_R$ proceeds analogously to Eqs.\eqref{81} and \eqref{82}, with the eigenbasis~\eqref{symlm}, in which the roles of the quantum numbers $m$ and $n$ are exchanged. 

Substituting the mode expansion into the equations~(\ref{111}) and (\ref{222}), using the eigenvalue equations~\eqref{tLlmn}, we obtain a coupled matrix equation whose characteristic equation yields the quartic equation
\begin{equation}
 E^4 - a  E^2 + b E + c = 0,
 \label{91}
\end{equation}
where the coefficients $a$, $b$, $c$ are given by 
\begin{equation}
\begin{aligned}
  a =\, &  2\left( M^2 +  \tilde{G}^2(\alpha,\beta_+)+   \tilde{\omega}^2_{m l} (\alpha, \beta_+)  + \frac14 e^{- 2\alpha} \Omega_3^2 \right), \\
  b =\, &  4 m \, e^{- 2 \alpha + 2 \beta_+} \tilde{G}(\alpha,\beta_+)  \Omega_3, \\
  c =\, &  \left( \frac{a}{2} - \frac12 e^{- 2\alpha} \Omega_3^2 \right)^2 + e^{- 4 \alpha - 2 \beta_+}\alpha^2_{m+\frac12}  \Omega^2_3\\
  &- 4 \, \tilde{\omega}^2_{m l} (\alpha, \beta_+) \tilde{G}^2(\alpha, \beta_+). 
  \label{93}
\end{aligned} 
\end{equation}
The solutions of this quartic yield the energy eigenvalues:
\begin{equation}
\begin{aligned}
  E^{s}_1 &= -\mathcal{A} \mp \mathcal{B}_+, \\
  E^{s}_2 &=  \mathcal{A} \pm \mathcal{B}_- ,
\end{aligned}
\qquad (s=1,2)
\end{equation}
where the upper (lower) sign corresponds to $s=1$ ($s=2$), and the auxiliary functions 
$\mathcal{A}$, $\mathcal{B}_{\pm}$ are defined as (with the expressions under the square roots assumed positive)
\begin{equation}
\begin{aligned}
&\mathcal{A}(\alpha, \beta_+, \Omega_{\hat{3}}): = \frac12 \sqrt{\frac{2 a}{3} + q_1 + q_2}, \\
&\mathcal{B}_{\pm}(\alpha, \beta_+, \Omega_{\hat{3}}):= \frac12 \sqrt{\frac{4 a}{3} - q_1 - q_2 \pm \frac{b}{\mathcal{A}}}, 
\label{95}
\end{aligned}
\end{equation}
where \(q_1, q_2\) are the roots of the cubic resolvent associated with the quartic in Eq.~\eqref{91}. 

The eigenvalues $E_1^{s}< 0$ correspond to particles, while $E_2^{s}> 0$ correspond to antiparticles. Specifically, the eigenvalues \( E_1^{s=1} \) are always negative and \( E_2^{s=1} \) are always positive. For $s=2$, the eigenvalues \( E^{s=2}_1 \) and \( E^{s=2}_2 \) can change sign if $\mathcal{B}_{-}(\alpha, \beta_+, \Omega_{\hat{3}}) > \mathcal{A} (\alpha, \beta_+, \Omega_{\hat{3}})$; however, they do so simultaneously, thus preserving charge conservation in curved spacetime.

In the rotating axisymmetric Bianchi IX background, the energy spectrum is modified by two contributions: the spin–dependent shifts due to anisotropy, and the rotational contribution from the spin–angular velocity coupling. The dependence on $\Omega_{\hat{3}}$ enters through all coefficients of the quartic equation~\eqref{91}, but the term proportional to $b$ plays a special role, as it enters with opposite signs in the functions $\mathcal{B}_{+}(\alpha, \beta+, \Omega_{\hat{3}})$ and $\mathcal{B}_{-}(\alpha, \beta+, \Omega_{\hat{3}})$. This sign difference leads to positive contributions for particle eigenstates and negative contributions for antiparticle eigenstates. Consequently, the spectrum splits into four distinct levels, corresponding to particles and antiparticles with two spin states each. An instructive analogy may be drawn with the Zeeman effect in quantum mechanics, where atomic energy levels split under an external magnetic field due to the interaction between the field and the electron’s magnetic moment. In the present context, the cosmic angular velocity plays the role of the magnetic field, producing a splitting of energy levels—a phenomenon absent in isotropic models. 

In the non-rotating limit $(\Omega_{\hat{3}} = 0)$, one has $b=0$ in~\eqref{93}, which yields $\mathcal{B}_{+}(\alpha, \beta_+, \Omega_{\hat{3}}) = \mathcal{B}_{-}(\alpha, \beta_+, \Omega_{\hat{3}})$; hence, the rotational asymmetry between particle and antiparticle eigenvalues disappears, and the spectrum reduces to the purely anisotropy-induced spin splitting discussed previously.

\section{Conclusion} \label{con}

In this work, we presented a geometric approach to addressing key cosmological puzzles of the standard $\Lambda$CDM model, focusing on the matter–antimatter asymmetry of the Universe. We have considered the Bianchi IX cosmological framework as a natural generalization of the isotropic FLRW geometry, allowing for both anisotropy and global rotation. This choice is supported by its relevance to early-universe dynamics via the BKL conjecture, its potential role in generating cosmological angular momentum, and studies suggesting that the axisymmetric Bianchi IX case may provide a framework for addressing certain CMB anomalies \cite{paper:Axisym}. Within this setting, we have explored whether the geometric structure of spacetime itself can induce asymmetries between particles and antiparticles, independent of extensions beyond the Standard Model.

We formulated the Dirac field in the Bianchi IX geometry using a Lagrangian approach, deriving the corresponding Dirac equation. The resulting equation contains coupling terms from geometric anisotropy and rotation, absent in FLRW spacetimes, which induce modifications in particle and antiparticle energy spectra. We further developed the harmonic analysis of spinor fields in the Bianchi IX background, extending earlier work on the Mixmaster universe \cite{paper:Dowker}, and solved the Dirac equations for non-rotating and rotating axisymmetric Bianchi IX models under a fixed-background approximation. 

The results highlight two key novelties relative to isotropic FLRW models, where the Dirac field has been studied previously~\cite{paper:ParkerS}. In the axisymmetric Bianchi IX model, spatial anisotropy induces spin-dependent modifications of the Dirac energy spectrum, with one spin state enhanced and the other suppressed for both particles and antiparticles. In the rotating axisymmetric case, in addition to these anisotropy-induced effects, rotational contributions from spin–angular velocity coupling appear with opposite signs for particles and antiparticles. These features demonstrate that spacetime geometry alone can imprint nontrivial structure on particle spectra. This motivates further studies of anisotropic geometric effects on creation and annihilation processes in QED and their potential implications for matter–antimatter asymmetry. While the observed asymmetry is conventionally associated with later cosmological epochs \cite{book:Kolb}, the early-universe geometric effects studied here provide initial conditions that can seed and influence the net asymmetry observed today.

Although this work focused on the axisymmetric subclass of Bianchi IX spacetimes, which captures the essential effects of anisotropy and rotation, preliminary analysis in~\cite{Vardanyan2025thesis} suggests that these qualitative features likely extend to the general Bianchi IX geometry, requiring a full treatment with more elaborate calculations. Furthermore, a natural next step is to extend the analysis to time-dependent Bianchi IX backgrounds using the adiabatic approximation, with further refinements within the WKB framework, to test the robustness of the spectral features under realistic cosmological conditions. Assessing the impact of geometric anisotropies on baryon asymmetry will require incorporating QCD in Bianchi IX backgrounds and examining sphaleron processes and CP-violating effects within the Standard Model. If geometric anisotropies amplify such effects beyond FLRW predictions, they may provide a natural, geometry-based explanation for the observed matter–antimatter asymmetry, highlighting a previously underappreciated role of spacetime geometry without invoking physics beyond the Standard Model.

\begin{acknowledgments}
I would like to thank Prof.\ Dr.\ Kiefer for valuable discussions and constructive feedback during the development of this work.
\end{acknowledgments}

\appendix

\section{Unitary irreducible representations of the group SU(2)}\label{irr}

Consider a unitary representation $g \rightarrow T_g$ of the group $\mathrm{SU(2)}$, which double-covers the rotation group $\mathrm{SO(3)}$,  acting on a Hilbert space $\mathcal{H}$. Each group element in a given representation can be expressed as
\begin{equation}
   T_g = e^{i \theta_k J_k },
\end{equation}
where $\theta_k$ are the continuous parameters and $J_k$ are the Hermitian generators of the representation. These generators satisfy the commutation relations
\begin{equation}
    [J_i, J_j] = i \varepsilon_{ijk} J_k,
    \label{JJJ}
\end{equation}
defining the Lie algebra \( \mathfrak{su}(2) \), which is isomorphic to \( \mathfrak{so}(3) \). To construct an eigenbasis of the generators, the ladder operators 
$ J_\pm = J_1 \pm i J_2$ are defined, which obey the commutation relations
\begin{equation}
 [J_+,J_3] = -J_+, \quad [J_-,J_3] = J_- , \quad [J_+,J_-] = 2 J_3.
\end{equation}
In any irreducible representation, $J_3$ has a discrete spectrum with eigenvectors forming an orthonormal basis, and $J_\pm$ act as raising and lowering operators within this basis. Moreover, all the eigenvectors of an irreducible representation with a given weight $l$ (with $l$ being either an integer or a half-integer) are also eigenstates of the Casimir operator $J^2= J_1^2 + J_2^2 + J_3^2$, with eigenvalues $l(l+1)$. A general representation decomposes into a direct sum of such irreducible components \cite{book:Gel}.

We begin by examining functions $f(\phi, \theta,\psi)$ on $\mathrm{SU}(2)$, which transform under the action of the group. The corresponding generators can be expressed in terms of the left- and right-invariant basis vector fields, as defined in Eqs.~(\ref{eright}) and (\ref{lr}):
\begin{equation}
L_i = i e_i, \quad \textnormal{and} \quad \tilde{L}_i = - i \tilde{e}_i,
\label{JRL}
\end{equation}
which satisfy the commutation relations in Eq.~\eqref{JJJ}. These correspond to the body-fixed and space-fixed angular momenta, respectively (see, e.g., \cite{book:Kemble}). Hence, the operators $L_\pm$, $L_3$ and $\tilde{L}_\pm$, $\tilde{L}_3$ take the following explicit forms:
\begin{equation}
\begin{aligned}
    &L_+ =  i e^{- i \psi} \left\{ i \frac{\partial}{\partial \theta} +  \cot{\theta} \frac{\partial}{\partial \psi}  - \frac{1}{\sin{\theta}} \frac{\partial}{\partial \phi} \right\},\\
    &L_- =   i e^{ i \psi} \left\{ - i \frac{\partial}{\partial \theta} + \cot{\theta} \frac{\partial}{\partial \psi}  - \frac{1}{\sin{\theta}} \frac{\partial}{\partial \phi} \right\}, \\
    & L_3 =  i \frac{\partial}{\partial \psi},
   \label{LLLD} 
\end{aligned} 
\end{equation}
and
\begin{equation}
\begin{aligned}
     &\tilde{L}_+ = i e^{i \phi} \left\{ - i \frac{\partial}{\partial \theta}  + \cot{\theta} \frac{\partial}{\partial \phi}  - \frac{1}{\sin{\theta}} \frac{\partial}{\partial \psi}   \right\},\\
     &\tilde{L}_- = i e^{- i \phi} \left\{i \frac{\partial}{\partial \theta}    +  \cot{\theta} \frac{\partial}{\partial \phi}  - \frac{1}{\sin{\theta}} \frac{\partial}{\partial \psi}  \right\},\\ 
     &\tilde{L}_3 = - i \frac{\partial}{\partial \phi}.
       \label{LLLTD}  
\end{aligned}  
\end{equation}
Additionally, the eigenvalue equation for the Casimir operator, $L^2 = \tilde{L}^2$, is given by 
\begin{equation}
\begin{aligned}
  \Bigg[\ & \frac{\partial^2}{\partial \theta^2} + \cot{\theta} \frac{\partial}{\partial \theta}  + \frac{1}{ \sin^2{\theta}} \left( \frac{\partial^2}{\partial \phi^2}   - 2 {\cos{\theta}}\frac{\partial}{\partial \phi}  \frac{\partial}{\partial \psi} + \frac{\partial^2}{\partial \psi^2}\right) \\
  &+ l(l + 1)\Bigg]f(\phi, \theta,\psi) = 0.  
\end{aligned}
\label{Casimir}
\end{equation}
The solutions to this equation are the matrix elements of the Wigner \( D^l \)-matrix\footnote{We adopt the \( zyz \)-convention for Euler angles, commonly used in quantum mechanics, which ensures that the Wigner \( D^l \)-matrix elements are real-valued.}, \( D^l_{-n m}(\phi, \theta, \psi) \) (see, e.g., \cite{book:Wigner, paper:Goldberg}), where \( |m| \leq l \) and \( |n| \leq l \). These functions, referred to as the \textit{generalized spherical functions}, form a complete orthonormal basis, satisfying the following orthonormality and completeness relations:
\begin{equation}
\begin{aligned}
&\int_0^{2 \pi} d \phi  \int_0^\pi \sin{\theta}  d \theta \int_0^{2 \pi} d \psi \, \Bar{D}^l_{-n m} (\phi, \theta, \psi) D^{l^\prime}_{- n^\prime m^\prime}(\phi, \theta, \psi)\\
&= \frac{8 \pi^2}{2 l + 1} \delta_{l l^\prime} \delta_{m m^\prime} \delta_{n n^\prime},
\label{orthD}  
\end{aligned}
\end{equation}
and
\begin{equation}
\begin{aligned}
&\sum_{l m n} \Bar{D}^l_{- n m}(\phi, \theta, \psi) D^{l}_{- n  m}(\phi^\prime, \theta^\prime, \psi^\prime)\\
&= \frac{8 \pi^2}{2 l + 1} \delta(\phi - \phi^\prime) \delta(\cos{\theta} - \cos{\theta^\prime}) \delta(\psi - \psi^\prime).
\label{normD}   
\end{aligned}
\end{equation}
For convenience, we denote the normalized eigenfunctions by
\begin{equation}
|l \, m \,  n \big> = \sqrt{\frac{2l+1}{8 \pi^2}}D^l_{- n \,  m }(\phi, \theta, \psi).
\label{orthket}
\end{equation}
The eigenvalue equations for the operators in (\ref{LLLD}) and (\ref{LLLTD}) then take the form
\begin{equation}
\begin{aligned}
& L_{+} |l \, m \,  n \rangle = \alpha_{n + 1}\, |l, \, m, \,  n + 1 \rangle, \\
& L_{-} |l \, m \, n \rangle = \alpha_n \, |l,\,  m, \, n - 1 \rangle,\\
& L_3 |l \, m \,  n \rangle = n |l \, m \, n \rangle,
\label{Llmn}
\end{aligned}
\end{equation}
and
\begin{equation}
\begin{aligned}
& \tilde{L}_{+} |l\, m \, n \rangle =  \alpha_{m + 1}|l,\, m + 1,\, n \rangle,\\
& \tilde{L}_{-} |l \, m \, n \rangle = \alpha_{m} |l, \, m - 1,\, n\rangle,\\
& \tilde{L}_3 |l \, m \, n \rangle = m |l \, m \, n \rangle,
\label{tLlmn}
\end{aligned}
\end{equation}
where $\alpha_x = \sqrt{(l+x)(l - x + 1)}$ with $x = m, n$. The quantum number $n$ is the eigenvalue of the operator $L_3$, and $m$ is the eigenvalue of $\tilde{L}_3$. The eigenvalue equation for the Casimir operator is
\begin{equation}
    L^2 |l\, m \, n \rangle = l(l+1) |l\, m \, n\rangle.
\end{equation}
Next, for the spinor representation, the generators are $j_i=\frac{\sigma_i}{2}$, where $\sigma_i$ are the Pauli matrices. The spinor eigenstates are 
\begin{equation}
  \Big| \frac12, \, \frac12 \Big\rangle = \left( \begin{matrix}
        1\\
       0
    \end{matrix}\right) \quad    \textnormal{and}  \quad \Big| \frac12, \,- \frac12 \Big\rangle  = \left( \begin{matrix}
       0\\
       1
    \end{matrix}\right).
   \label{speig} 
\end{equation}
We denote the spinor eigenstates as $| s \, n_s \big>$, with the weight of $s=\frac12$ and $n_s = -\frac12, \frac12$.

Finally, for the tensor product of the function representation with weight $l$ and the spinor representation with weight $s=\frac12$, one possible choice of eigenbasis is 
\begin{equation}
 | l s; m \, (n - n_s) \, n_s \rangle = |l , m ,  n - n_s \rangle |s \, n_s \rangle ,
 \label{A15}
\end{equation}
which forms an orthonormal system of eigenvectors of $J_3$ with eigenvalues $n$, where $-l -  \frac12 \leq n \leq l + \frac12$. Additionally, the product can also be decomposed into a direct sum of irreducible representations of weight $j$, where $j= l \pm \frac12$. These form a canonical basis denoted by $\{ | l s; j\, m \,  n\rangle \}$, which can be expressed as linear combinations of the basis $\{|l , m ,  n - n_s \rangle |s \, n_s \rangle \}$ \cite{book:Gel}:
\begin{equation}
  | l s; j\, m \,  n\rangle = \sum  C^{j n}_{ l,n - n_s;s n_s}|l , m ,  n - n_s\rangle |s \, n_s \rangle ,
\label{A17}
\end{equation}
where $j = l\pm \frac12$ and $C^{j n}_{ l, n - n_s;s n_s}$ are the Clebsch-Gordan coefficients. The components of the Clebsch-Gordan coefficients are given in the Table~\ref{table:species5} below: 
\begin{table}[ht]
\centering 
\begin{tabular}{  c | c  c   } 
$s  = \frac12$ & $n_s  = \frac12$ & $n_s  = -\frac12$  \\ [0.5ex] 
\hline 
 $j = l + \frac12$ &	$ \sqrt{\frac{l + m + \frac12}{2 l + 1}}$ &	$ \sqrt{\frac{l - m + \frac12}{2 l + 1}}$ \\ [1ex]
$j = l - \frac12$ &	$ - \sqrt{\frac{l - m + \frac12}{2 l + 1}}$ &	$ \sqrt{\frac{l + m + \frac12}{2 l + 1}}$ \\ [1ex] 
\end{tabular}
\caption{The Clebsch-Gordan coefficients for $s = \frac12$ \cite{book:Sakurai}.}
\label{table:species5} 
\end{table}

Thus, the eigenvectors take the explicit form
\begin{equation}
\begin{aligned}
    \Big| l \, \frac12; j\, m \,  n \Big\rangle 
  = \frac{1}{\sqrt{2 l + 1}} \left( \begin{matrix}
        \pm \sqrt{ l \pm n + \frac12} \, |l , m ,n - \frac12\rangle \\ 
      \sqrt{ l \mp n + \frac12}  \,|l , m ,n + \frac12\rangle \\ 
    \end{matrix} \right),
    \label{jnmg}  
\end{aligned}
\end{equation}
with $j = l \pm \frac12$. These generalize the spin spherical harmonics \cite{book:Ed}; we therefore refer to them as the \textit{generalized spin spherical harmonics}.


\section{Eigenstates of standard and ideal asymmetric tops} \label{Atop}

\subsection{Rotational eigenstates of rigid tops} \label{rottop}

The quantum rotational eigenstates of rigid tops are reviewed here, providing the necessary background for the construction of the ideal symmetric top Hamiltonian eigenstates in Sec.~\ref{HA}, as well as the ideal asymmetric top Hamiltonian eigenstates discussed in the following section.

The Hamiltonian of a rigid rotor is given by
\begin{equation}
H = a L_1^2 + b L_2^2 + c L_3^2,
\label{B1}
\end{equation}
where $L_1$, $L_2$, $L_3$ are the body-fixed angular momentum components defined in Eq.~\eqref{JRL}. The rotational constants are
\begin{equation}
a = \frac{1}{2 I_1}, \quad b = \frac{1}{2 I_2}, \quad c = \frac{1}{2 I_3},
\end{equation}
with $I_1$, $I_2$, and $I_3$ denoting the principal moments of inertia.

For a symmetric top with $I_1 = I_2 \neq I_3$ (prolate configuration), the rotational constants satisfy $a = b$, and the Hamiltonian becomes
\begin{equation}
H = a L^2 + (c - a) L_3^2.
\label{HamB}
\end{equation}
In the special case of a spherical top, all principal moments of inertia are equal, $I_1 = I_2 = I_3$. The Hamiltonian then simplifies to $H = a L^2$, and the eigenvalue problem coincides with that of the Casimir operator $L^2$ in Eq.~\eqref{Casimir}. The eigenvalues are $E = a l (l + 1)$ with corresponding eigenfunctions given by the orthonormal set $|l \, m \, n \rangle$ defined in Eq.~\eqref{orthket}, where $l$ takes integer values and $|n|, |m|\leq l$.

Furthermore, using Eqs.~\eqref{LLLD} and \eqref{Casimir}, the symmetric top eigenvalue equation for the Hamiltonian~\eqref{HamB}  takes the explicit form
\begin{equation}
\begin{aligned}
 - a \Bigg\{ \frac{\partial^2}{\partial \theta^2} & + \cot{\theta} \frac{\partial}{\partial \theta} + \left(\frac{c}{a} + \cot^2{\theta} \right) \frac{\partial^2}{\partial \psi^2}+ \frac{1}{\sin^2{\theta}}\frac{\partial^2}{\partial \phi^2} \\
 & - \frac{2 \cos{\theta}}{\sin^2{\theta}}\frac{\partial^2}{\partial \phi \partial \psi}\Bigg\} \chi(\phi, \theta,\psi) = E \chi(\phi, \theta,\psi). 
\end{aligned}
 \label{symL}
\end{equation}
The eigenfunctions of this equation are the same as those of the spherical top \cite{book:Ed}, which is expected as the Hamiltonian of the symmetric top~\eqref{HamB} commutes with the operators $L^2$, $L_3$, and $\tilde{L}_3$. The corresponding eigenvalues are $E = a l (l + 1) + (c - a) n^2$.

In an asymmetric top, all three principal moments of inertia are different, $I_1 \neq I_2 \neq I_3$ in the Hamiltonin~\eqref{B1}. In this case the Hamiltonian commutes with $L^2$ and $\tilde{L}_3$, but not with $L_3$. As a result, $n$ is not a good quantum number, and the eigenstates can be written as linear combinations of the symmetric-top eigenstates $|l \, m \,  n \big>$~\cite{book:Dav}:
\begin{equation}
    |l \, m \big> = \sum_{n = - l}^l a_n |l \, m \,  n \big>,
    \label{lma}
\end{equation}
where $|l \, m  \big>$ denotes an eigenstate of the asymmetric top. Inserting this expansion into the eigenvalue equation of Hamiltonian~\eqref{B1}, gives a set of coupled equations for the coefficients $a_n$,
\begin{equation}
\sum_{n}  a_n  \left(  H_{n n^\prime}  -   E \delta_{n n^\prime} \right) = 0,
\end{equation}
with the Hamiltonian matrix elements defined by
\begin{equation}
\hat{H} |l \, m \,  n \big> = \sum_{n^\prime} H_{n n^\prime} |l \, m \,  n^\prime \big>.
\end{equation}
For a given $l$, this leads to $(2l+1)$ coupled equations for each value of $n$. The energy eigenvalues of the asymmetric top are the roots of the secular equation
\begin{equation}
   \textnormal{det} |H_{n n^\prime} - E \delta_{ n n^\prime}| = 0.
\end{equation}
The calculation of asymmetric-top energy levels can be simplified by choosing an eigenbasis that incorporates the symmetry properties of the top~\cite{paper:Winter, paper:CKing, paper:Mull}. The symmetry group of the asymmetric top is the dihedral group $D_2$, which consists of the identity $E$ and three two-fold rotations $C_2^a$, $C_2^b$, and $C_2^c$ about the three principal axes by angle $\pi$. The Euler angles $\{\phi, \theta,\psi\}$ transform under the operations $C_2^a$, $C_2^b$, $C_2^c$ as expressed in Table \ref{table:EulerC} below:
\begin{table}[ht]
\centering 
\begin{tabular}{  c | c | c  } 
$C_2^a$ &	$C_2^b$  &	$C_2^c$ \\ [0.5ex] 
\hline 
$- \phi$&	$\phi$&	$\phi$ \\ [1ex]
$\theta + \pi$ & $\theta + \pi $&	$\theta$  \\ [1ex] 
$\psi$ &	$- \psi$ &	$\psi + \pi$ \\ [1ex] 
\end{tabular}
  \caption{Transformation of the Euler angles under the operation of two-fold rotations.}
\label{table:EulerC} 
\end{table}

Both the Hamiltonian~\eqref{B1} and the angular momentum commutation relations~\eqref{JJJ} are invariant under these symmetry operations. Consequently, the eigenfunctions can be classified according to the irreducible representations of the $D_2$ group. This group has four one-dimensional irreducible representations: $A$, $B_1$, $B_2$, and $B_3$ \cite{book:Landau3}. The $A$ representation is symmetric, while the $B$ representations are antisymmetric with respect to the symmetry operations of the group elements, as summarized in Table~\ref{table:species2}.

\begin{table}[ht]
\centering 
\begin{tabular}{  c | c  c  c  c  } 
${\mathbf{D}}_2$ &  $E$	& $C_2^a$ &	$C_2^b$  &	$C_2^c$ \\ [0.5ex] 
\hline 
 $A$ &	$1$ &	$1$&	$1$&	$1$ \\ [1ex]
$B_a$ &	$1$ &	$1$ & $-1$&	$ - 1$  \\ [1ex] 
$B_b$ & $1$ &	$-1$ &	$1$ &	$- 1$ \\ [1ex] 
$B_c$ & $1$ &	$ - 1$ &	$-1$ &	$1$ \\ [1ex] 
\end{tabular}
  \caption{Characters of irreducible representations of $D_2$ group.}
\label{table:species2} 
\end{table}

The symmetric–top eigenstates $|l \,  m \, n \big>$ transform under the symmetry group operations as
\begin{equation}
\begin{aligned}
   C_2^a |l \, m \,n \big>  &= (-1)^{l} |l\,  m\, - n \big>, \\
   C_2^b |l \, m \, n \big>  &= (-1)^{l + n }  |l\,  m \, - n \big>,\\
   C_2^c |l \, m \, n \big>  &= (-1)^n |l \,  m \,  n \big>.
   \label{ccclmn}
\end{aligned}
\end{equation}
Following Ref.~\cite{paper:VV}, we introduce for $n \neq 0$ the functions
\begin{equation}
   |l \, m  \, n \, x\big> = (- 1)^{max \ n,m} | l\, m \, n\big>,
\end{equation}
and form the Wang combinations 
\begin{equation}
|l \, m \, n \, \gamma\big> = 2^{ -1/2} \left[ |l\,  m \, n \, x\big> + (- 1)^\gamma |l \, m\,  - n\, x\big> \right],
\label{Wang}
\end{equation}
where $\gamma = 0$ (even) or $1$ (odd). For $n = 0$, only the even combination exists,
\begin{equation}
|l\, m \, 0 \,0\big> =  |l \, m \, 0 \,x\big>.
\end{equation}
These states transform as
\begin{equation}
\begin{aligned}
   C_2^a |l\, m\, n\, \gamma\big> &= (-1)^{l + \gamma} |l\, m \, n \, \gamma\big>,  \\
   C_2^b |l \, m \, n \, \gamma\big> &= (-1)^{l+ \gamma + n} |l\, m \, n\, \gamma\big>, \\
   C_2^c |l \, m \, n \, \gamma\big> &= (-1)^{n} |l \, m \, n \, \gamma\big>. 
\end{aligned}
\end{equation}
Thus, $|l \, m \, n \, \gamma\big>$ are simultaneous eigenstates of 
$C_2^a$, $C_2^b$, and $C_2^c$. Their symmetry species (or the representation classes) depend on the parities of $n$ and $l + \gamma$. 

With this eigenbasis, the asymmetric–top eigenstates can be expanded as
\begin{equation}
  |l \, m \,  \gamma\big> = \sum_{n} a^l_{m n, \gamma}  |l\, m \, n \, \gamma\big>,
\end{equation}
and the secular equation becomes
\begin{equation}
    \textnormal{det} |H_{n n^\prime, \gamma} - E \delta_{ n n^\prime}| = 0.
\end{equation}
By expressing the asymmetric top eigenstates in the $|l , m , n , \gamma\big>$ eigenbasis, the secular equation naturally decomposes into four lower-degree equations, each associated with one of the symmetry species (irreducible representations) of the eigenstates.

\subsection{Eigenstates of the ideal asymmetric top} \label{aTop}

The body-fixed Hamiltonian of the ideal asymmetric top, given by Eq.~(\ref{asit}), commutes with the operators \( L^2 = \tilde{L}^2 \), \( j^2 \), \( J^2 \), and \( \tilde{L}_3 \). However, in contrast to the ideal symmetric top discussed in Sec.~\ref{HAs}, it does not commute with $L_3$, $j_3$, or $J_3$, implying that $n$ is no longer a good quantum number. Consequently, as demonstrated in \cite{paper:Dowker}, the eigenbasis of the ideal asymmetric top—analogous to that of the standard asymmetric top—can be expressed as a linear combination of the symmetric-top eigenbasis \cite{book:Dav}.

To be able to discuss the spinor components separately, we introduce a spinor index $\mu$, which takes values $\mu = 1,2$. The eigenstates of the ideal asymmetric top for each spinor component, denoted by $| l \, s;  m \rangle_\mu$, can be expanded in terms of the ideal symmetric top eigenstates~\eqref{symlm} as
\begin{equation}
 | l \, s;  m \rangle_\mu = \sum_n  a^l_{m n} \Big| l, m, n + \frac{(-1)^\mu}{2} \Big\rangle_\mu,
    \label{rightasym}
\end{equation}
where $a^l_{m n}$ are the expansion coefficients. Alternatively, when using the ideal symmetric top eigenbasis~\eqref{symjn} (here abbreviated as $|j \, m \, n \rangle$ for convenience, instead of the full notation $| l \,s ; j \, m\, n \rangle$), the eigenstates take the form:
\begin{equation}
  | j\, m  \rangle_\mu = \sum_n  a^j_{m n} | j\,  m \, n  \rangle_\mu,
\end{equation}
where
\begin{equation}
   | j \, m \,n \rangle_\mu  
   =  \sum_{j = l\pm \frac12}  C^{ j, n}_{l, n + \frac{(-1)^\mu}{2}; \frac12, \frac{(-1)^\mu}{2}} \Big| l, m, n + \frac{(-1)^\mu}{2} \Big\rangle_\mu.  
\end{equation}
Thus, just as in the symmetric case, two equivalent but distinct eigenbases exist for the ideal asymmetric top.

The space-fixed Hamiltonian (\ref{asitB}) commutes with the operators
${\tilde{J}}^2$, $L^2 = {\tilde{L}}^2$, ${\tilde{j}}^2$, and $L_3$, but does not commute with $\tilde{J}_3$, $\tilde{L}_3$, or $\tilde{j}_3$. Therefore, a summation over the quantum number $m$ is required. Accordingly, for the ideal asymmetric top eigenbasis for each spinor component admits the expansion via  the basis (\ref{symln}) as 
\begin{equation}
  | l \, s;  n \rangle_\mu = \sum_m  \tilde{a}^l_{m n} \Big| l, m + \frac{(-1)^\mu}{2}, n  \Big\rangle_\mu. 
 \label{leftasym}
\end{equation}
Similarly, for the eigenbasis given in (\ref{symjm}), the eigenstates can be expressed as
\begin{equation}
  | j\, n \rangle_\mu = \sum_m  \tilde{a}^j_{m n} | j\,  m \, n  \rangle_\mu,
\end{equation}
where
\begin{equation}
  | j \, m \,n \rangle_\mu  
   =  \sum_{j = l\pm \frac12}  C^{ j, m}_{l, m + \frac{(-1)^\mu}{2}; \frac12, \frac{(-1)^\mu}{2}} \Big| l,  m + \frac{(-1)^\mu}{2}, n \Big\rangle_\mu.  
\end{equation}
Finally, one can incorporate the ideal asymmetric top’s symmetry properties to expand its eigenstates in terms of the symmetry-adapted basis, generalizing the approach used for the standard asymmetric top. To implement this, under the action of the operators $C_2^a$, $C_2^b$ and $C_2^c$, the eigenstates $|j\,  m \, n  \rangle$ are transformed as 
\begin{equation}
\begin{aligned}
   C_2^a |j\,  m \, n  \rangle &= e^{i \pi j} |j\, m \, - n \rangle, \\
   C_2^b |j \, m \, n \rangle &= e^{i \pi (j + n)} |j \, m \, - n \rangle,\\
   C_2^c |j \, m\,  n  \rangle  &=  e^{i \pi n} |j \,  m \,  n \rangle.
\label{cccjmn}
\end{aligned}
\end{equation}
We then introduce a symmetry-adapted basis
\begin{equation}
|j\,  m \,  n \, \gamma\rangle = 2^{ -1/2} \left[ |j\, m \, n \, x\rangle + (- 1)^\gamma |j \, m \, - n\,  x \rangle \right],
\label{Wangj}
\end{equation}
which generalizes the standard asymmetric-top basis~(\ref{Wang}). Further details on these states, their transformation under the symmetry group, and the resulting block decomposition of the Hamiltonian can be found in \cite{paper:Dowker}. The eigenstates of the ideal asymmetric top Hamiltonian~\eqref{asit} can then be expanded as
\begin{equation}
  | j\, m \,\gamma \rangle_\mu = \sum_n a^j_{mn,\gamma}  | j\, m\, n \,\gamma \rangle_\mu.
\end{equation}
For the space-fixed Hamiltonian~\eqref{asitB}, corresponding to the general Bianchi IX case, the expansion is analogous and can be obtained from the above relation by interchanging the quantum numbers $m$ and $n$. Let us also point out, however, that this expansion is not convenient for solving the Dirac equation, because the operators $L_i$ and $\tilde{L}_i$ appear linearly in these equations and do not commute with the $D_2$ symmetry operators $C_2^a$, $C_2^b$, and $C_2^c$.

\bibliography{refs}

\begin{thebibliography}{91}%
\makeatletter
\providecommand \@ifxundefined [1]{%
 \@ifx{#1\undefined}
}%
\providecommand \@ifnum [1]{%
 \ifnum #1\expandafter \@firstoftwo
 \else \expandafter \@secondoftwo
 \fi
}%
\providecommand \@ifx [1]{%
 \ifx #1\expandafter \@firstoftwo
 \else \expandafter \@secondoftwo
 \fi
}%
\providecommand \natexlab [1]{#1}%
\providecommand \enquote  [1]{``#1''}%
\providecommand \bibnamefont  [1]{#1}%
\providecommand \bibfnamefont [1]{#1}%
\providecommand \citenamefont [1]{#1}%
\providecommand \href@noop [0]{\@secondoftwo}%
\providecommand \href [0]{\begingroup \@sanitize@url \@href}%
\providecommand \@href[1]{\@@startlink{#1}\@@href}%
\providecommand \@@href[1]{\endgroup#1\@@endlink}%
\providecommand \@sanitize@url [0]{\catcode `\\12\catcode `\$12\catcode `\&12\catcode `\#12\catcode `\^12\catcode `\_12\catcode `\%12\relax}%
\providecommand \@@startlink[1]{}%
\providecommand \@@endlink[0]{}%
\providecommand \url  [0]{\begingroup\@sanitize@url \@url }%
\providecommand \@url [1]{\endgroup\@href {#1}{\urlprefix }}%
\providecommand \urlprefix  [0]{URL }%
\providecommand \Eprint [0]{\href }%
\providecommand \doibase [0]{http://dx.doi.org/}%
\providecommand \selectlanguage [0]{\@gobble}%
\providecommand \bibinfo  [0]{\@secondoftwo}%
\providecommand \bibfield  [0]{\@secondoftwo}%
\providecommand \translation [1]{[#1]}%
\providecommand \BibitemOpen [0]{}%
\providecommand \bibitemStop [0]{}%
\providecommand \bibitemNoStop [0]{.\EOS\space}%
\providecommand \EOS [0]{\spacefactor3000\relax}%
\providecommand \BibitemShut  [1]{\csname bibitem#1\endcsname}%
\let\auto@bib@innerbib\@empty
\bibitem [{\citenamefont {Bennett}\ \emph {et~al.}(2003)\citenamefont {Bennett} \emph {et~al.}}]{paper:Wmap}%
  \BibitemOpen
  \bibfield  {author} {\bibinfo {author} {\bibfnamefont {C.}~\bibnamefont {Bennett}} \emph {et~al.},\ }\href@noop {} {\bibfield  {journal} {\bibinfo  {journal} {Astrophys. J.}\ }\textbf {\bibinfo {volume} {583}},\ \bibinfo {pages} {1} (\bibinfo {year} {2003})},\ \Eprint {http://arxiv.org/abs/astro-ph/0301158} {arXiv:astro-ph/0301158} \BibitemShut {NoStop}%
\bibitem [{\citenamefont {{Planck Collaboration}}(2020)}]{paper:Planck}%
  \BibitemOpen
  \bibfield  {author} {\bibinfo {author} {\bibnamefont {{Planck Collaboration}}},\ }\href@noop {} {\bibfield  {journal} {\bibinfo  {journal} {Astron. Astrophys.}\ }\textbf {\bibinfo {volume} {641}},\ \bibinfo {pages} {A6} (\bibinfo {year} {2020})},\ \bibinfo {note} {erratum: Astron. Astrophys. \textbf{652}, C4 (2021)},\ \Eprint {http://arxiv.org/abs/1807.06209} {arXiv:1807.06209} \BibitemShut {NoStop}%
\bibitem [{\citenamefont {Adelman-McCarthy}\ \emph {et~al.}(2006)\citenamefont {Adelman-McCarthy} \emph {et~al.}}]{paper:Ad}%
  \BibitemOpen
  \bibfield  {author} {\bibinfo {author} {\bibfnamefont {J.~K.}\ \bibnamefont {Adelman-McCarthy}} \emph {et~al.},\ }\href@noop {} {\bibfield  {journal} {\bibinfo  {journal} {Astrophys. J. Suppl.}\ }\textbf {\bibinfo {volume} {162}},\ \bibinfo {pages} {38} (\bibinfo {year} {2006})},\ \Eprint {http://arxiv.org/abs/astro-ph/0507711} {arXiv:astro-ph/0507711} \BibitemShut {NoStop}%
\bibitem [{\citenamefont {Perivolaropoulos}\ and\ \citenamefont {Skara}(2022)}]{paper:Greek}%
  \BibitemOpen
  \bibfield  {author} {\bibinfo {author} {\bibfnamefont {L.}~\bibnamefont {Perivolaropoulos}}\ and\ \bibinfo {author} {\bibfnamefont {F.}~\bibnamefont {Skara}},\ }\href@noop {} {\bibfield  {journal} {\bibinfo  {journal} {New Astron. Rev.}\ }\textbf {\bibinfo {volume} {95}},\ \bibinfo {pages} {101659} (\bibinfo {year} {2022})}\BibitemShut {NoStop}%
\bibitem [{\citenamefont {Popolo}\ and\ \citenamefont {Delliou}(2017)}]{paper:smallscalep}%
  \BibitemOpen
  \bibfield  {author} {\bibinfo {author} {\bibfnamefont {A.~D.}\ \bibnamefont {Popolo}}\ and\ \bibinfo {author} {\bibfnamefont {M.~L.}\ \bibnamefont {Delliou}},\ }\href@noop {} {\bibfield  {journal} {\bibinfo  {journal} {Galaxies}\ }\textbf {\bibinfo {volume} {5}},\ \bibinfo {pages} {17} (\bibinfo {year} {2017})},\ \Eprint {http://arxiv.org/abs/1606.07790v3} {arXiv:1606.07790v3} \BibitemShut {NoStop}%
\bibitem [{\citenamefont {Aluri}\ \emph {et~al.}(2023)\citenamefont {Aluri} \emph {et~al.}}]{Aluri:2022hzs}%
  \BibitemOpen
  \bibfield  {author} {\bibinfo {author} {\bibfnamefont {P.~K.}\ \bibnamefont {Aluri}} \emph {et~al.},\ }\href {\doibase 10.1088/1361-6382/acbefc} {\bibfield  {journal} {\bibinfo  {journal} {Class. Quant. Grav.}\ }\textbf {\bibinfo {volume} {40}},\ \bibinfo {pages} {094001} (\bibinfo {year} {2023})},\ \Eprint {http://arxiv.org/abs/2207.05765} {arXiv:2207.05765 [astro-ph.CO]} \BibitemShut {NoStop}%
\bibitem [{\citenamefont {Belinskii}\ \emph {et~al.}(1970)\citenamefont {Belinskii}, \citenamefont {Khalatnikov},\ and\ \citenamefont {Lifshitz}}]{paper:BKL1}%
  \BibitemOpen
  \bibfield  {author} {\bibinfo {author} {\bibfnamefont {V.~A.}\ \bibnamefont {Belinskii}}, \bibinfo {author} {\bibfnamefont {I.~M.}\ \bibnamefont {Khalatnikov}}, \ and\ \bibinfo {author} {\bibfnamefont {E.~M.}\ \bibnamefont {Lifshitz}},\ }\href@noop {} {\bibfield  {journal} {\bibinfo  {journal} {Adv. Phys.}\ }\textbf {\bibinfo {volume} {19}},\ \bibinfo {pages} {525} (\bibinfo {year} {1970})}\BibitemShut {NoStop}%
\bibitem [{\citenamefont {Belinskii}\ \emph {et~al.}(1982)\citenamefont {Belinskii}, \citenamefont {Khalatnikov},\ and\ \citenamefont {Lifshitz}}]{paper:BKL2}%
  \BibitemOpen
  \bibfield  {author} {\bibinfo {author} {\bibfnamefont {V.~A.}\ \bibnamefont {Belinskii}}, \bibinfo {author} {\bibfnamefont {I.~M.}\ \bibnamefont {Khalatnikov}}, \ and\ \bibinfo {author} {\bibfnamefont {E.~M.}\ \bibnamefont {Lifshitz}},\ }\href@noop {} {\bibfield  {journal} {\bibinfo  {journal} {Adv. Phys.}\ }\textbf {\bibinfo {volume} {31}},\ \bibinfo {pages} {639} (\bibinfo {year} {1982})}\BibitemShut {NoStop}%
\bibitem [{\citenamefont {Li}(1998)}]{paper:Li}%
  \BibitemOpen
  \bibfield  {author} {\bibinfo {author} {\bibfnamefont {L.-X.}\ \bibnamefont {Li}},\ }\href@noop {} {\bibfield  {journal} {\bibinfo  {journal} {Gen. Rel. Grav.}\ }\textbf {\bibinfo {volume} {30}},\ \bibinfo {pages} {497} (\bibinfo {year} {1998})}\BibitemShut {NoStop}%
\bibitem [{\citenamefont {Ciufolini}\ and\ \citenamefont {Wheeler}(1995)}]{book:CW}%
  \BibitemOpen
  \bibfield  {author} {\bibinfo {author} {\bibfnamefont {I.}~\bibnamefont {Ciufolini}}\ and\ \bibinfo {author} {\bibfnamefont {J.~A.}\ \bibnamefont {Wheeler}},\ }\href@noop {} {\emph {\bibinfo {title} {Gravitation and Inertia}}}\ (\bibinfo  {publisher} {Princeton University Press},\ \bibinfo {address} {Princeton},\ \bibinfo {year} {1995})\BibitemShut {NoStop}%
\bibitem [{\citenamefont {Gibbons}(1979)}]{paper:Gib1}%
  \BibitemOpen
  \bibfield  {author} {\bibinfo {author} {\bibfnamefont {G.~W.}\ \bibnamefont {Gibbons}},\ }\href@noop {} {\bibfield  {journal} {\bibinfo  {journal} {Phys. Lett. B}\ }\textbf {\bibinfo {volume} {84}},\ \bibinfo {pages} {431} (\bibinfo {year} {1979})}\BibitemShut {NoStop}%
\bibitem [{\citenamefont {Gibbons}(1980)}]{paper:Gib3}%
  \BibitemOpen
  \bibfield  {author} {\bibinfo {author} {\bibfnamefont {G.~W.}\ \bibnamefont {Gibbons}},\ }\href@noop {} {\bibfield  {journal} {\bibinfo  {journal} {Ann. Phys.}\ }\textbf {\bibinfo {volume} {125}},\ \bibinfo {pages} {98} (\bibinfo {year} {1980})}\BibitemShut {NoStop}%
\bibitem [{\citenamefont {Obregon}\ and\ \citenamefont {Ryan}(1981)}]{paper:ORyan}%
  \BibitemOpen
  \bibfield  {author} {\bibinfo {author} {\bibfnamefont {O.}~\bibnamefont {Obregon}}\ and\ \bibinfo {author} {\bibfnamefont {M.}~\bibnamefont {Ryan}},\ }\href@noop {} {\bibfield  {journal} {\bibinfo  {journal} {J. Math. Phys.}\ }\textbf {\bibinfo {volume} {22}},\ \bibinfo {pages} {623} (\bibinfo {year} {1981})}\BibitemShut {NoStop}%
\bibitem [{\citenamefont {Damour}\ and\ \citenamefont {Spindel}(2011)}]{paper:Damour}%
  \BibitemOpen
  \bibfield  {author} {\bibinfo {author} {\bibfnamefont {T.}~\bibnamefont {Damour}}\ and\ \bibinfo {author} {\bibfnamefont {P.}~\bibnamefont {Spindel}},\ }\href@noop {} {\bibfield  {journal} {\bibinfo  {journal} {Phys. Rev. D}\ }\textbf {\bibinfo {volume} {83}},\ \bibinfo {pages} {123520} (\bibinfo {year} {2011})}\BibitemShut {NoStop}%
\bibitem [{\citenamefont {Sundell}\ and\ \citenamefont {Koivisto}(2015)}]{paper:Axisym}%
  \BibitemOpen
  \bibfield  {author} {\bibinfo {author} {\bibfnamefont {P.}~\bibnamefont {Sundell}}\ and\ \bibinfo {author} {\bibfnamefont {T.}~\bibnamefont {Koivisto}},\ }\href@noop {} {\bibfield  {journal} {\bibinfo  {journal} {Phys. Rev. D}\ }\textbf {\bibinfo {volume} {92}},\ \bibinfo {pages} {123529} (\bibinfo {year} {2015})}\BibitemShut {NoStop}%
\bibitem [{\citenamefont {Birrell}\ and\ \citenamefont {Davies}(1982)}]{book:Birrell}%
  \BibitemOpen
  \bibfield  {author} {\bibinfo {author} {\bibfnamefont {N.~D.}\ \bibnamefont {Birrell}}\ and\ \bibinfo {author} {\bibfnamefont {P.~C.~W.}\ \bibnamefont {Davies}},\ }\href@noop {} {\emph {\bibinfo {title} {Quantum Fields in Curved Space}}}\ (\bibinfo  {publisher} {Cambridge University Press},\ \bibinfo {address} {Cambridge},\ \bibinfo {year} {1982})\BibitemShut {NoStop}%
\bibitem [{\citenamefont {Vardanyan}(2025)}]{Vardanyan2025thesis}%
  \BibitemOpen
  \bibfield  {author} {\bibinfo {author} {\bibfnamefont {T.}~\bibnamefont {Vardanyan}},\ }\emph {\bibinfo {title} {Exploring Matter-Antimatter Asymmetry in a Rotating Universe}},\ \href@noop {} {\bibinfo {type} {Ph.d. thesis}},\ \bibinfo  {school} {University of Cologne} (\bibinfo {year} {2025}),\ \bibinfo {note} {\url{https://kups.ub.uni-koeln.de/78502/}}\BibitemShut {NoStop}%
\bibitem [{\citenamefont {Dowker}\ and\ \citenamefont {Pettengill}(1974)}]{paper:Dowker}%
  \BibitemOpen
  \bibfield  {author} {\bibinfo {author} {\bibfnamefont {J.~S.}\ \bibnamefont {Dowker}}\ and\ \bibinfo {author} {\bibfnamefont {D.~F.}\ \bibnamefont {Pettengill}},\ }\href@noop {} {\bibfield  {journal} {\bibinfo  {journal} {J. Phys. A}\ }\textbf {\bibinfo {volume} {7}},\ \bibinfo {pages} {1527} (\bibinfo {year} {1974})}\BibitemShut {NoStop}%
\bibitem [{\citenamefont {Hu}(1973)}]{paper:HuI}%
  \BibitemOpen
  \bibfield  {author} {\bibinfo {author} {\bibfnamefont {B.~L.}\ \bibnamefont {Hu}},\ }\href@noop {} {\bibfield  {journal} {\bibinfo  {journal} {Phys. Rev. D}\ }\textbf {\bibinfo {volume} {8}},\ \bibinfo {pages} {1048} (\bibinfo {year} {1973})}\BibitemShut {NoStop}%
\bibitem [{\citenamefont {Mather}\ \emph {et~al.}(1990)\citenamefont {Mather} \emph {et~al.}}]{paper:Cobe}%
  \BibitemOpen
  \bibfield  {author} {\bibinfo {author} {\bibfnamefont {J.~C.}\ \bibnamefont {Mather}} \emph {et~al.},\ }\href@noop {} {\bibfield  {journal} {\bibinfo  {journal} {Astrophys. J. Lett.}\ }\textbf {\bibinfo {volume} {354}},\ \bibinfo {pages} {L37} (\bibinfo {year} {1990})}\BibitemShut {NoStop}%
\bibitem [{\citenamefont {Copi}\ \emph {et~al.}(2010)\citenamefont {Copi}, \citenamefont {Huterer}, \citenamefont {Schwarz},\ and\ \citenamefont {Starkman}}]{paper:Dom}%
  \BibitemOpen
  \bibfield  {author} {\bibinfo {author} {\bibfnamefont {C.~J.}\ \bibnamefont {Copi}}, \bibinfo {author} {\bibfnamefont {D.}~\bibnamefont {Huterer}}, \bibinfo {author} {\bibfnamefont {D.~J.}\ \bibnamefont {Schwarz}}, \ and\ \bibinfo {author} {\bibfnamefont {G.~D.}\ \bibnamefont {Starkman}},\ }\href@noop {} {\bibfield  {journal} {\bibinfo  {journal} {Adv. Astron.}\ }\textbf {\bibinfo {volume} {2010}},\ \bibinfo {pages} {847541} (\bibinfo {year} {2010})},\ \Eprint {http://arxiv.org/abs/1004.5602v2} {arXiv:1004.5602v2} \BibitemShut {NoStop}%
\bibitem [{\citenamefont {Schwarz}\ \emph {et~al.}(2016)\citenamefont {Schwarz}, \citenamefont {Copi}, \citenamefont {Huterer},\ and\ \citenamefont {Starkman}}]{paper:CMBa}%
  \BibitemOpen
  \bibfield  {author} {\bibinfo {author} {\bibfnamefont {D.~J.}\ \bibnamefont {Schwarz}}, \bibinfo {author} {\bibfnamefont {C.~J.}\ \bibnamefont {Copi}}, \bibinfo {author} {\bibfnamefont {D.}~\bibnamefont {Huterer}}, \ and\ \bibinfo {author} {\bibfnamefont {G.~D.}\ \bibnamefont {Starkman}},\ }\href@noop {} {\bibfield  {journal} {\bibinfo  {journal} {Class. Quant. Grav.}\ }\textbf {\bibinfo {volume} {33}},\ \bibinfo {pages} {184001} (\bibinfo {year} {2016})}\BibitemShut {NoStop}%
\bibitem [{\citenamefont {Bennett}\ \emph {et~al.}(2011)\citenamefont {Bennett}, \citenamefont {Hill}, \citenamefont {Hinshaw}, \citenamefont {Larson}, \citenamefont {Smith}, \citenamefont {Dunkley} \emph {et~al.}}]{paper:stat}%
  \BibitemOpen
  \bibfield  {author} {\bibinfo {author} {\bibfnamefont {C.~L.}\ \bibnamefont {Bennett}}, \bibinfo {author} {\bibfnamefont {R.~S.}\ \bibnamefont {Hill}}, \bibinfo {author} {\bibfnamefont {G.}~\bibnamefont {Hinshaw}}, \bibinfo {author} {\bibfnamefont {D.}~\bibnamefont {Larson}}, \bibinfo {author} {\bibfnamefont {K.~M.}\ \bibnamefont {Smith}}, \bibinfo {author} {\bibfnamefont {J.}~\bibnamefont {Dunkley}},  \emph {et~al.},\ }\href@noop {} {\bibfield  {journal} {\bibinfo  {journal} {Astrophys. J. Supp.}\ }\textbf {\bibinfo {volume} {192}},\ \bibinfo {pages} {17} (\bibinfo {year} {2011})},\ \Eprint {http://arxiv.org/abs/1001.4758v2} {arXiv:1001.4758v2} \BibitemShut {NoStop}%
\bibitem [{\citenamefont {Abdalla}\ \emph {et~al.}(2022)\citenamefont {Abdalla} \emph {et~al.}}]{paper:ab}%
  \BibitemOpen
  \bibfield  {author} {\bibinfo {author} {\bibfnamefont {E.}~\bibnamefont {Abdalla}} \emph {et~al.},\ }\href@noop {} {\bibfield  {journal} {\bibinfo  {journal} {JHEAp}\ }\textbf {\bibinfo {volume} {34}},\ \bibinfo {pages} {49} (\bibinfo {year} {2022})}\BibitemShut {NoStop}%
\bibitem [{\citenamefont {Famaey}\ and\ \citenamefont {McGaugh}(2012)}]{paper:Fam}%
  \BibitemOpen
  \bibfield  {author} {\bibinfo {author} {\bibfnamefont {B.}~\bibnamefont {Famaey}}\ and\ \bibinfo {author} {\bibfnamefont {S.}~\bibnamefont {McGaugh}},\ }\href@noop {} {\bibfield  {journal} {\bibinfo  {journal} {Living Rev. Rel.}\ }\textbf {\bibinfo {volume} {15}},\ \bibinfo {pages} {10} (\bibinfo {year} {2012})}\BibitemShut {NoStop}%
\bibitem [{\citenamefont {Blas}(2019)}]{notes:Blas}%
  \BibitemOpen
  \bibfield  {author} {\bibinfo {author} {\bibfnamefont {D.}~\bibnamefont {Blas}},\ }\href@noop {} {\enquote {\bibinfo {title} {Introduction to dark matter},}\ }\bibinfo {howpublished} {Durham University} (\bibinfo {year} {2019}),\ \bibinfo {note} {available at \url{https://conference.ippp.dur.ac.uk/event/1291/attachments/6076/8176/DM_Blas1.pdf}}\BibitemShut {NoStop}%
\bibitem [{\citenamefont {Freese}(2009)}]{paper:Kat}%
  \BibitemOpen
  \bibfield  {author} {\bibinfo {author} {\bibfnamefont {K.}~\bibnamefont {Freese}},\ }\href@noop {} {\bibfield  {journal} {\bibinfo  {journal} {EAS Publ. Ser.}\ }\textbf {\bibinfo {volume} {36}},\ \bibinfo {pages} {113} (\bibinfo {year} {2009})},\ \Eprint {http://arxiv.org/abs/0812.4005} {arXiv:0812.4005} \BibitemShut {NoStop}%
\bibitem [{\citenamefont {Bergström}(2000)}]{paper:Lars}%
  \BibitemOpen
  \bibfield  {author} {\bibinfo {author} {\bibfnamefont {L.}~\bibnamefont {Bergström}},\ }\href@noop {} {\bibfield  {journal} {\bibinfo  {journal} {Rept. Prog. Phys.}\ }\textbf {\bibinfo {volume} {63}},\ \bibinfo {pages} {793} (\bibinfo {year} {2000})}\BibitemShut {NoStop}%
\bibitem [{\citenamefont {Bertone}\ \emph {et~al.}(2005)\citenamefont {Bertone}, \citenamefont {Hooper},\ and\ \citenamefont {Silk}}]{paper:Silk}%
  \BibitemOpen
  \bibfield  {author} {\bibinfo {author} {\bibfnamefont {G.}~\bibnamefont {Bertone}}, \bibinfo {author} {\bibfnamefont {D.}~\bibnamefont {Hooper}}, \ and\ \bibinfo {author} {\bibfnamefont {J.}~\bibnamefont {Silk}},\ }\href@noop {} {\bibfield  {journal} {\bibinfo  {journal} {Phys. Rept.}\ }\textbf {\bibinfo {volume} {405}},\ \bibinfo {pages} {279} (\bibinfo {year} {2005})}\BibitemShut {NoStop}%
\bibitem [{\citenamefont {van~den Bosch}\ \emph {et~al.}(2001)\citenamefont {van~den Bosch}, \citenamefont {Burkert},\ and\ \citenamefont {Swaters}}]{paper:Bosch}%
  \BibitemOpen
  \bibfield  {author} {\bibinfo {author} {\bibfnamefont {F.~C.}\ \bibnamefont {van~den Bosch}}, \bibinfo {author} {\bibfnamefont {A.}~\bibnamefont {Burkert}}, \ and\ \bibinfo {author} {\bibfnamefont {R.~A.}\ \bibnamefont {Swaters}},\ }\href@noop {} {\bibfield  {journal} {\bibinfo  {journal} {Mon. Not. Roy. Astron. Soc.}\ }\textbf {\bibinfo {volume} {326}},\ \bibinfo {pages} {1205} (\bibinfo {year} {2001})}\BibitemShut {NoStop}%
\bibitem [{\citenamefont {Viel}\ \emph {et~al.}(2013)\citenamefont {Viel}, \citenamefont {Becker}, \citenamefont {Bolton},\ and\ \citenamefont {Haehnelt}}]{paper:Matt}%
  \BibitemOpen
  \bibfield  {author} {\bibinfo {author} {\bibfnamefont {M.}~\bibnamefont {Viel}}, \bibinfo {author} {\bibfnamefont {G.~D.}\ \bibnamefont {Becker}}, \bibinfo {author} {\bibfnamefont {J.~S.}\ \bibnamefont {Bolton}}, \ and\ \bibinfo {author} {\bibfnamefont {M.~G.}\ \bibnamefont {Haehnelt}},\ }\href@noop {} {\bibfield  {journal} {\bibinfo  {journal} {Phys. Rev. D}\ }\textbf {\bibinfo {volume} {88}},\ \bibinfo {pages} {043502} (\bibinfo {year} {2013})}\BibitemShut {NoStop}%
\bibitem [{\citenamefont {Bullock}\ and\ \citenamefont {Boylan-Kolchin}(2017)}]{paper:ssp}%
  \BibitemOpen
  \bibfield  {author} {\bibinfo {author} {\bibfnamefont {J.~S.}\ \bibnamefont {Bullock}}\ and\ \bibinfo {author} {\bibfnamefont {M.}~\bibnamefont {Boylan-Kolchin}},\ }\href@noop {} {\bibfield  {journal} {\bibinfo  {journal} {Ann. Rev. Astron. Astrophys.}\ }\textbf {\bibinfo {volume} {55}},\ \bibinfo {pages} {343} (\bibinfo {year} {2017})},\ \Eprint {http://arxiv.org/abs/1707.04256v2} {arXiv:1707.04256v2} \BibitemShut {NoStop}%
\bibitem [{\citenamefont {Hu}\ \emph {et~al.}(2000)\citenamefont {Hu}, \citenamefont {Barkana},\ and\ \citenamefont {Gruzinov}}]{paper:Fuzzy}%
  \BibitemOpen
  \bibfield  {author} {\bibinfo {author} {\bibfnamefont {W.}~\bibnamefont {Hu}}, \bibinfo {author} {\bibfnamefont {R.}~\bibnamefont {Barkana}}, \ and\ \bibinfo {author} {\bibfnamefont {A.}~\bibnamefont {Gruzinov}},\ }\href@noop {} {\bibfield  {journal} {\bibinfo  {journal} {Phys. Rev. Lett.}\ }\textbf {\bibinfo {volume} {85}},\ \bibinfo {pages} {1158} (\bibinfo {year} {2000})}\BibitemShut {NoStop}%
\bibitem [{\citenamefont {Steigman}(1976)}]{paper:SS}%
  \BibitemOpen
  \bibfield  {author} {\bibinfo {author} {\bibfnamefont {G.}~\bibnamefont {Steigman}},\ }\href@noop {} {\bibfield  {journal} {\bibinfo  {journal} {Ann. Rev. Astron. Astrophys.}\ }\textbf {\bibinfo {volume} {14}},\ \bibinfo {pages} {339} (\bibinfo {year} {1976})}\BibitemShut {NoStop}%
\bibitem [{\citenamefont {Steigman}(2008)}]{paper:antiU}%
  \BibitemOpen
  \bibfield  {author} {\bibinfo {author} {\bibfnamefont {G.}~\bibnamefont {Steigman}},\ }\href@noop {} {\bibfield  {journal} {\bibinfo  {journal} {JCAP}\ }\textbf {\bibinfo {volume} {0810}},\ \bibinfo {pages} {001} (\bibinfo {year} {2008})},\ \Eprint {http://arxiv.org/abs/0808.1122} {arXiv:0808.1122} \BibitemShut {NoStop}%
\bibitem [{\citenamefont {Rubakov}\ and\ \citenamefont {Gorbunov}(2017)}]{book:Rub}%
  \BibitemOpen
  \bibfield  {author} {\bibinfo {author} {\bibfnamefont {V.~A.}\ \bibnamefont {Rubakov}}\ and\ \bibinfo {author} {\bibfnamefont {D.~S.}\ \bibnamefont {Gorbunov}},\ }\href@noop {} {\emph {\bibinfo {title} {Introduction to the Theory of the Early Universe: Hot Big Bang Theory}}},\ \bibinfo {edition} {2nd}\ ed.\ (\bibinfo  {publisher} {World Scientific},\ \bibinfo {address} {Singapore},\ \bibinfo {year} {2017})\BibitemShut {NoStop}%
\bibitem [{\citenamefont {Canetti}\ \emph {et~al.}(2012)\citenamefont {Canetti}, \citenamefont {Drewes},\ and\ \citenamefont {Shaposhnikov}}]{paper:maas}%
  \BibitemOpen
  \bibfield  {author} {\bibinfo {author} {\bibfnamefont {L.}~\bibnamefont {Canetti}}, \bibinfo {author} {\bibfnamefont {M.}~\bibnamefont {Drewes}}, \ and\ \bibinfo {author} {\bibfnamefont {M.}~\bibnamefont {Shaposhnikov}},\ }\href@noop {} {\bibfield  {journal} {\bibinfo  {journal} {New J. Phys.}\ }\textbf {\bibinfo {volume} {14}},\ \bibinfo {pages} {095012} (\bibinfo {year} {2012})},\ \Eprint {http://arxiv.org/abs/1204.4186v2} {arXiv:1204.4186v2} \BibitemShut {NoStop}%
\bibitem [{\citenamefont {Steigman}(2010)}]{paper:etaBBN}%
  \BibitemOpen
  \bibfield  {author} {\bibinfo {author} {\bibfnamefont {G.}~\bibnamefont {Steigman}},\ }\href@noop {} {\bibfield  {journal} {\bibinfo  {journal} {PoS NICXI}\ ,\ \bibinfo {pages} {001}} (\bibinfo {year} {2010})},\ \Eprint {http://arxiv.org/abs/1008.4765} {arXiv:1008.4765} \BibitemShut {NoStop}%
\bibitem [{\citenamefont {et~al. (WMAP)}(2011)}]{paper:etaCMB}%
  \BibitemOpen
  \bibfield  {author} {\bibinfo {author} {\bibfnamefont {E.~K.}\ \bibnamefont {et~al. (WMAP)}},\ }\href@noop {} {\bibfield  {journal} {\bibinfo  {journal} {Astrophys. J. Suppl.}\ }\textbf {\bibinfo {volume} {192}},\ \bibinfo {pages} {18} (\bibinfo {year} {2011})},\ \Eprint {http://arxiv.org/abs/1001.4538} {arXiv:1001.4538} \BibitemShut {NoStop}%
\bibitem [{\citenamefont {Sakharov}(1967)}]{paper:Sakharov}%
  \BibitemOpen
  \bibfield  {author} {\bibinfo {author} {\bibfnamefont {A.~D.}\ \bibnamefont {Sakharov}},\ }\href@noop {} {\bibfield  {journal} {\bibinfo  {journal} {JETP Lett.}\ }\textbf {\bibinfo {volume} {6}},\ \bibinfo {pages} {24} (\bibinfo {year} {1967})}\BibitemShut {NoStop}%
\bibitem [{\citenamefont {Kuzmin}\ \emph {et~al.}(1985)\citenamefont {Kuzmin}, \citenamefont {Rubakov},\ and\ \citenamefont {Shaposhnikov}}]{paper:Kuzmin}%
  \BibitemOpen
  \bibfield  {author} {\bibinfo {author} {\bibfnamefont {V.~A.}\ \bibnamefont {Kuzmin}}, \bibinfo {author} {\bibfnamefont {V.~A.}\ \bibnamefont {Rubakov}}, \ and\ \bibinfo {author} {\bibfnamefont {M.~E.}\ \bibnamefont {Shaposhnikov}},\ }\href@noop {} {\bibfield  {journal} {\bibinfo  {journal} {Phys. Lett. B}\ }\textbf {\bibinfo {volume} {155}},\ \bibinfo {pages} {36} (\bibinfo {year} {1985})}\BibitemShut {NoStop}%
\bibitem [{\citenamefont {Davoudiasl}\ \emph {et~al.}(2004)\citenamefont {Davoudiasl}, \citenamefont {Kitano}, \citenamefont {Kribs}, \citenamefont {Murayama},\ and\ \citenamefont {Steinhardt}}]{paper:GB}%
  \BibitemOpen
  \bibfield  {author} {\bibinfo {author} {\bibfnamefont {H.}~\bibnamefont {Davoudiasl}}, \bibinfo {author} {\bibfnamefont {R.}~\bibnamefont {Kitano}}, \bibinfo {author} {\bibfnamefont {G.~D.}\ \bibnamefont {Kribs}}, \bibinfo {author} {\bibfnamefont {H.}~\bibnamefont {Murayama}}, \ and\ \bibinfo {author} {\bibfnamefont {P.~J.}\ \bibnamefont {Steinhardt}},\ }\href@noop {} {\bibfield  {journal} {\bibinfo  {journal} {Phys. Rev. Lett.}\ }\textbf {\bibinfo {volume} {93}},\ \bibinfo {pages} {201301} (\bibinfo {year} {2004})},\ \Eprint {http://arxiv.org/abs/hep-ph/0403019} {arXiv:hep-ph/0403019} \BibitemShut {NoStop}%
\bibitem [{\citenamefont {Dine}\ and\ \citenamefont {Kusenko}(2003)}]{paper:Dine}%
  \BibitemOpen
  \bibfield  {author} {\bibinfo {author} {\bibfnamefont {M.}~\bibnamefont {Dine}}\ and\ \bibinfo {author} {\bibfnamefont {A.}~\bibnamefont {Kusenko}},\ }\href@noop {} {\bibfield  {journal} {\bibinfo  {journal} {Rev. Mod. Phys.}\ }\textbf {\bibinfo {volume} {76}},\ \bibinfo {pages} {1} (\bibinfo {year} {2003})}\BibitemShut {NoStop}%
\bibitem [{\citenamefont {Rubakov}\ and\ \citenamefont {Shaposhnikov}(1996)}]{paper:Rub}%
  \BibitemOpen
  \bibfield  {author} {\bibinfo {author} {\bibfnamefont {V.~A.}\ \bibnamefont {Rubakov}}\ and\ \bibinfo {author} {\bibfnamefont {M.~E.}\ \bibnamefont {Shaposhnikov}},\ }\href@noop {} {\bibfield  {journal} {\bibinfo  {journal} {Usp. Fiz. Nauk}\ }\textbf {\bibinfo {volume} {166}},\ \bibinfo {pages} {493} (\bibinfo {year} {1996})}\BibitemShut {NoStop}%
\bibitem [{\citenamefont {Riotto}\ and\ \citenamefont {Trodden}(1999)}]{paper:Trod}%
  \BibitemOpen
  \bibfield  {author} {\bibinfo {author} {\bibfnamefont {A.}~\bibnamefont {Riotto}}\ and\ \bibinfo {author} {\bibfnamefont {M.}~\bibnamefont {Trodden}},\ }\href@noop {} {\bibfield  {journal} {\bibinfo  {journal} {Ann. Rev. Nucl. Part. Sci.}\ }\textbf {\bibinfo {volume} {49}},\ \bibinfo {pages} {35} (\bibinfo {year} {1999})}\BibitemShut {NoStop}%
\bibitem [{\citenamefont {Rubakov}\ and\ \citenamefont {Gorbunov}(2011)}]{book:Rubakov}%
  \BibitemOpen
  \bibfield  {author} {\bibinfo {author} {\bibfnamefont {V.~A.}\ \bibnamefont {Rubakov}}\ and\ \bibinfo {author} {\bibfnamefont {D.~S.}\ \bibnamefont {Gorbunov}},\ }\href@noop {} {\emph {\bibinfo {title} {Introduction to the Theory of the Early Universe: Cosmological Perturbations and Inflationary Theory}}}\ (\bibinfo  {publisher} {World Scientific},\ \bibinfo {address} {Singapore},\ \bibinfo {year} {2011})\BibitemShut {NoStop}%
\bibitem [{\citenamefont {Ellis}\ \emph {et~al.}(2012)\citenamefont {Ellis}, \citenamefont {Maartens},\ and\ \citenamefont {MacCallum}}]{book:EMM}%
  \BibitemOpen
  \bibfield  {author} {\bibinfo {author} {\bibfnamefont {G.~F.~R.}\ \bibnamefont {Ellis}}, \bibinfo {author} {\bibfnamefont {R.}~\bibnamefont {Maartens}}, \ and\ \bibinfo {author} {\bibfnamefont {M.~A.~H.}\ \bibnamefont {MacCallum}},\ }\href@noop {} {\emph {\bibinfo {title} {Relativistic Cosmology}}}\ (\bibinfo  {publisher} {Cambridge University Press},\ \bibinfo {address} {Cambridge},\ \bibinfo {year} {2012})\BibitemShut {NoStop}%
\bibitem [{\citenamefont {Grib}\ \emph {et~al.}(1994)\citenamefont {Grib}, \citenamefont {Mamaev},\ and\ \citenamefont {Mostepanenko}}]{book:Grib}%
  \BibitemOpen
  \bibfield  {author} {\bibinfo {author} {\bibfnamefont {A.~A.}\ \bibnamefont {Grib}}, \bibinfo {author} {\bibfnamefont {S.~G.}\ \bibnamefont {Mamaev}}, \ and\ \bibinfo {author} {\bibfnamefont {V.~M.}\ \bibnamefont {Mostepanenko}},\ }\href@noop {} {\emph {\bibinfo {title} {Vacuum Quantum Eﬀects in Strong Fields}}}\ (\bibinfo  {publisher} {Friedmann Laboratory Publishing},\ \bibinfo {address} {St. Petersburg},\ \bibinfo {year} {1994})\BibitemShut {NoStop}%
\bibitem [{\citenamefont {Landau}\ and\ \citenamefont {Lifshitz}(1975)}]{book:LandauF}%
  \BibitemOpen
  \bibfield  {author} {\bibinfo {author} {\bibfnamefont {L.~D.}\ \bibnamefont {Landau}}\ and\ \bibinfo {author} {\bibfnamefont {E.~M.}\ \bibnamefont {Lifshitz}},\ }\href@noop {} {\emph {\bibinfo {title} {The Classical Theory of Fields}}},\ \bibinfo {edition} {4th}\ ed.\ (\bibinfo  {publisher} {Pergamon Press},\ \bibinfo {address} {Oxford},\ \bibinfo {year} {1975})\BibitemShut {NoStop}%
\bibitem [{\citenamefont {Lifshitz}(1946)}]{paper:Lif}%
  \BibitemOpen
  \bibfield  {author} {\bibinfo {author} {\bibfnamefont {E.}~\bibnamefont {Lifshitz}},\ }\href@noop {} {\bibfield  {journal} {\bibinfo  {journal} {J. Phys. (USSR)}\ }\textbf {\bibinfo {volume} {10}},\ \bibinfo {pages} {116} (\bibinfo {year} {1946})}\BibitemShut {NoStop}%
\bibitem [{\citenamefont {Misner}(1969)}]{paper:Misner69}%
  \BibitemOpen
  \bibfield  {author} {\bibinfo {author} {\bibfnamefont {C.~W.}\ \bibnamefont {Misner}},\ }\href@noop {} {\bibfield  {journal} {\bibinfo  {journal} {Phys. Rev. Lett.}\ }\textbf {\bibinfo {volume} {22}},\ \bibinfo {pages} {1071} (\bibinfo {year} {1969})}\BibitemShut {NoStop}%
\bibitem [{\citenamefont {Ryan}(1971{\natexlab{a}})}]{paper:Ryan1}%
  \BibitemOpen
  \bibfield  {author} {\bibinfo {author} {\bibfnamefont {M.~P.}\ \bibnamefont {Ryan}},\ }\href@noop {} {\bibfield  {journal} {\bibinfo  {journal} {Ann. Phys.}\ }\textbf {\bibinfo {volume} {65}},\ \bibinfo {pages} {506} (\bibinfo {year} {1971}{\natexlab{a}})}\BibitemShut {NoStop}%
\bibitem [{\citenamefont {Ryan}(1971{\natexlab{b}})}]{paper:Ryan2}%
  \BibitemOpen
  \bibfield  {author} {\bibinfo {author} {\bibfnamefont {M.~P.}\ \bibnamefont {Ryan}},\ }\href@noop {} {\bibfield  {journal} {\bibinfo  {journal} {Ann. Phys.}\ }\textbf {\bibinfo {volume} {68}},\ \bibinfo {pages} {541} (\bibinfo {year} {1971}{\natexlab{b}})}\BibitemShut {NoStop}%
\bibitem [{\citenamefont {Hu}\ and\ \citenamefont {Parker}(1978)}]{paper:HuP}%
  \BibitemOpen
  \bibfield  {author} {\bibinfo {author} {\bibfnamefont {B.~L.}\ \bibnamefont {Hu}}\ and\ \bibinfo {author} {\bibfnamefont {L.}~\bibnamefont {Parker}},\ }\href@noop {} {\bibfield  {journal} {\bibinfo  {journal} {Phys. Rev. D}\ }\textbf {\bibinfo {volume} {17}},\ \bibinfo {pages} {933} (\bibinfo {year} {1978})}\BibitemShut {NoStop}%
\bibitem [{\citenamefont {van~den Hoogen}\ and\ \citenamefont {Olasagasti}(1999)}]{paper:Is}%
  \BibitemOpen
  \bibfield  {author} {\bibinfo {author} {\bibfnamefont {R.}~\bibnamefont {van~den Hoogen}}\ and\ \bibinfo {author} {\bibfnamefont {I.}~\bibnamefont {Olasagasti}},\ }\href@noop {} {\bibfield  {journal} {\bibinfo  {journal} {Phys. Rev. D}\ }\textbf {\bibinfo {volume} {59}},\ \bibinfo {pages} {107302} (\bibinfo {year} {1999})}\BibitemShut {NoStop}%
\bibitem [{\citenamefont {Gibbons}\ and\ \citenamefont {Hawking}(1977)}]{paper:GH}%
  \BibitemOpen
  \bibfield  {author} {\bibinfo {author} {\bibfnamefont {G.~W.}\ \bibnamefont {Gibbons}}\ and\ \bibinfo {author} {\bibfnamefont {S.~W.}\ \bibnamefont {Hawking}},\ }\href@noop {} {\bibfield  {journal} {\bibinfo  {journal} {Phys. Rev. D}\ }\textbf {\bibinfo {volume} {15}},\ \bibinfo {pages} {2738} (\bibinfo {year} {1977})}\BibitemShut {NoStop}%
\bibitem [{\citenamefont {Hawking}\ and\ \citenamefont {Moss}(1982)}]{paper:HM}%
  \BibitemOpen
  \bibfield  {author} {\bibinfo {author} {\bibfnamefont {S.~W.}\ \bibnamefont {Hawking}}\ and\ \bibinfo {author} {\bibfnamefont {I.~G.}\ \bibnamefont {Moss}},\ }\href@noop {} {\bibfield  {journal} {\bibinfo  {journal} {Phys. Lett.}\ }\textbf {\bibinfo {volume} {110B}},\ \bibinfo {pages} {35} (\bibinfo {year} {1982})}\BibitemShut {NoStop}%
\bibitem [{\citenamefont {Wald}(1983)}]{paper:Wald}%
  \BibitemOpen
  \bibfield  {author} {\bibinfo {author} {\bibfnamefont {R.~M.}\ \bibnamefont {Wald}},\ }\href@noop {} {\bibfield  {journal} {\bibinfo  {journal} {Phys. Rev. D}\ }\textbf {\bibinfo {volume} {28}},\ \bibinfo {pages} {2118} (\bibinfo {year} {1983})}\BibitemShut {NoStop}%
\bibitem [{\citenamefont {Peebles}(1980)}]{book:Peebles}%
  \BibitemOpen
  \bibfield  {author} {\bibinfo {author} {\bibfnamefont {P.~J.~E.}\ \bibnamefont {Peebles}},\ }\href@noop {} {\emph {\bibinfo {title} {The Large-Scale Structure of the Universe}}}\ (\bibinfo  {publisher} {Princeton University Press},\ \bibinfo {address} {Princeton},\ \bibinfo {year} {1980})\BibitemShut {NoStop}%
\bibitem [{\citenamefont {Gamow}(1946)}]{paper:Gamow}%
  \BibitemOpen
  \bibfield  {author} {\bibinfo {author} {\bibfnamefont {G.}~\bibnamefont {Gamow}},\ }\href@noop {} {\bibfield  {journal} {\bibinfo  {journal} {Nature}\ }\textbf {\bibinfo {volume} {158}},\ \bibinfo {pages} {549} (\bibinfo {year} {1946})}\BibitemShut {NoStop}%
\bibitem [{\citenamefont {Gödel}(1990)}]{paper:Godel1}%
  \BibitemOpen
  \bibfield  {author} {\bibinfo {author} {\bibfnamefont {K.}~\bibnamefont {Gödel}}\ }(\bibinfo  {publisher} {Oxford University Press},\ \bibinfo {address} {Oxford},\ \bibinfo {year} {1990})\BibitemShut {NoStop}%
\bibitem [{\citenamefont {Collins}\ and\ \citenamefont {Hawking}(1973)}]{paper:HawkC}%
  \BibitemOpen
  \bibfield  {author} {\bibinfo {author} {\bibfnamefont {C.~B.}\ \bibnamefont {Collins}}\ and\ \bibinfo {author} {\bibfnamefont {S.~W.}\ \bibnamefont {Hawking}},\ }\href@noop {} {\bibfield  {journal} {\bibinfo  {journal} {Mon. Not. R. Astron. Soc.}\ }\textbf {\bibinfo {volume} {162}},\ \bibinfo {pages} {307} (\bibinfo {year} {1973})}\BibitemShut {NoStop}%
\bibitem [{\citenamefont {Obukhov}(2000)}]{paper:Ob}%
  \BibitemOpen
  \bibfield  {author} {\bibinfo {author} {\bibfnamefont {Y.~N.}\ \bibnamefont {Obukhov}},\ }in\ \href@noop {} {\emph {\bibinfo {booktitle} {Colloquium on Cosmic Rotation}}},\ \bibinfo {editor} {edited by\ \bibinfo {editor} {\bibfnamefont {M.}~\bibnamefont {Scherfner}} \emph {et~al.}}\ (\bibinfo  {publisher} {Wissenschaft und Technik Verlag},\ \bibinfo {address} {Berlin},\ \bibinfo {year} {2000})\ pp.\ \bibinfo {pages} {23--96},\ \bibinfo {note} {arXiv:astro-ph/0008106}\BibitemShut {NoStop}%
\bibitem [{\citenamefont {Godłowski}\ \emph {et~al.}(2003)\citenamefont {Godłowski}, \citenamefont {Szydłowski}, \citenamefont {Flin},\ and\ \citenamefont {Biernacka}}]{paper:Godl2}%
  \BibitemOpen
  \bibfield  {author} {\bibinfo {author} {\bibfnamefont {W.}~\bibnamefont {Godłowski}}, \bibinfo {author} {\bibfnamefont {M.}~\bibnamefont {Szydłowski}}, \bibinfo {author} {\bibfnamefont {P.}~\bibnamefont {Flin}}, \ and\ \bibinfo {author} {\bibfnamefont {M.}~\bibnamefont {Biernacka}},\ }\href@noop {} {\bibfield  {journal} {\bibinfo  {journal} {Gen. Rel. and Grav.}\ }\textbf {\bibinfo {volume} {35}},\ \bibinfo {pages} {907} (\bibinfo {year} {2003})}\BibitemShut {NoStop}%
\bibitem [{\citenamefont {Godłowski}(2011)}]{paper:Godl11}%
  \BibitemOpen
  \bibfield  {author} {\bibinfo {author} {\bibfnamefont {W.}~\bibnamefont {Godłowski}},\ }\href@noop {} {\bibfield  {journal} {\bibinfo  {journal} {Int. J. Mod. Phys. D}\ }\textbf {\bibinfo {volume} {20}},\ \bibinfo {pages} {1643} (\bibinfo {year} {2011})}\BibitemShut {NoStop}%
\bibitem [{\citenamefont {Wang}\ \emph {et~al.}(2021)\citenamefont {Wang}, \citenamefont {Libeskind}, \citenamefont {Tempel}, \citenamefont {Kang},\ and\ \citenamefont {Guo}}]{paper:Wang}%
  \BibitemOpen
  \bibfield  {author} {\bibinfo {author} {\bibfnamefont {P.}~\bibnamefont {Wang}}, \bibinfo {author} {\bibfnamefont {N.~I.}\ \bibnamefont {Libeskind}}, \bibinfo {author} {\bibfnamefont {E.}~\bibnamefont {Tempel}}, \bibinfo {author} {\bibfnamefont {X.}~\bibnamefont {Kang}}, \ and\ \bibinfo {author} {\bibfnamefont {Q.}~\bibnamefont {Guo}},\ }\href@noop {} {\bibfield  {journal} {\bibinfo  {journal} {Nature Astron.}\ }\textbf {\bibinfo {volume} {5}},\ \bibinfo {pages} {839} (\bibinfo {year} {2021})}\BibitemShut {NoStop}%
\bibitem [{\citenamefont {Hwang}\ and\ \citenamefont {Lee}(2007)}]{paper:rotcl}%
  \BibitemOpen
  \bibfield  {author} {\bibinfo {author} {\bibfnamefont {H.~S.}\ \bibnamefont {Hwang}}\ and\ \bibinfo {author} {\bibfnamefont {M.~G.}\ \bibnamefont {Lee}},\ }\href@noop {} {\bibfield  {journal} {\bibinfo  {journal} {Astrophys. J.}\ }\textbf {\bibinfo {volume} {662}},\ \bibinfo {pages} {236} (\bibinfo {year} {2007})}\BibitemShut {NoStop}%
\bibitem [{\citenamefont {Kiefer}(2025)}]{book:Kiefer}%
  \BibitemOpen
  \bibfield  {author} {\bibinfo {author} {\bibfnamefont {C.}~\bibnamefont {Kiefer}},\ }\href@noop {} {\emph {\bibinfo {title} {Quantum Gravity}}},\ \bibinfo {edition} {4th}\ ed.\ (\bibinfo  {publisher} {Oxford University Press},\ \bibinfo {address} {Oxford},\ \bibinfo {year} {2025})\BibitemShut {NoStop}%
\bibitem [{\citenamefont {Jantzen}(1987)}]{paper:Jantzen}%
  \BibitemOpen
  \bibfield  {author} {\bibinfo {author} {\bibfnamefont {R.~T.}\ \bibnamefont {Jantzen}},\ }\bibfield  {booktitle} {\emph {\bibinfo {booktitle} {Proceedings of the International School Enrico Fermi, Course LXXXVI (1982) on Gamov Cosmology}},\ }\href@noop {} {\ ,\ \bibinfo {pages} {61} (\bibinfo {year} {1987})},\ \bibinfo {note} {arXiv:gr-qc/0102035}\BibitemShut {NoStop}%
\bibitem [{\citenamefont {Misner}(1968)}]{paper:Misner68}%
  \BibitemOpen
  \bibfield  {author} {\bibinfo {author} {\bibfnamefont {C.~W.}\ \bibnamefont {Misner}},\ }\href@noop {} {\bibfield  {journal} {\bibinfo  {journal} {Astrophys. J.}\ }\textbf {\bibinfo {volume} {151}},\ \bibinfo {pages} {431} (\bibinfo {year} {1968})}\BibitemShut {NoStop}%
\bibitem [{\citenamefont {Ryan}(1972)}]{book:R}%
  \BibitemOpen
  \bibfield  {author} {\bibinfo {author} {\bibfnamefont {M.~P.}\ \bibnamefont {Ryan}},\ }\href@noop {} {\emph {\bibinfo {title} {Hamiltonian Cosmology}}}\ (\bibinfo  {publisher} {Springer},\ \bibinfo {address} {Berlin},\ \bibinfo {year} {1972})\BibitemShut {NoStop}%
\bibitem [{\citenamefont {Ryan}\ and\ \citenamefont {Shepley}(1975)}]{book:RS}%
  \BibitemOpen
  \bibfield  {author} {\bibinfo {author} {\bibfnamefont {M.~P.}\ \bibnamefont {Ryan}}\ and\ \bibinfo {author} {\bibfnamefont {L.~C.}\ \bibnamefont {Shepley}},\ }\href@noop {} {\emph {\bibinfo {title} {Homogeneous Relativistic Cosmologies}}}\ (\bibinfo  {publisher} {Princeton University Press},\ \bibinfo {address} {Princeton},\ \bibinfo {year} {1975})\BibitemShut {NoStop}%
\bibitem [{\citenamefont {Jantzen}(1979)}]{paper:Jan}%
  \BibitemOpen
  \bibfield  {author} {\bibinfo {author} {\bibfnamefont {R.~T.}\ \bibnamefont {Jantzen}},\ }\href@noop {} {\bibfield  {journal} {\bibinfo  {journal} {Commun. math. Phys.}\ }\textbf {\bibinfo {volume} {64}},\ \bibinfo {pages} {211} (\bibinfo {year} {1979})}\BibitemShut {NoStop}%
\bibitem [{\citenamefont {Jantzen}\ and\ \citenamefont {Uggla}(1999)}]{paper:UJ}%
  \BibitemOpen
  \bibfield  {author} {\bibinfo {author} {\bibfnamefont {R.~T.}\ \bibnamefont {Jantzen}}\ and\ \bibinfo {author} {\bibfnamefont {C.}~\bibnamefont {Uggla}},\ }\href@noop {} {\bibfield  {journal} {\bibinfo  {journal} {J. Math. Phys.}\ }\textbf {\bibinfo {volume} {40}},\ \bibinfo {pages} {353} (\bibinfo {year} {1999})}\BibitemShut {NoStop}%
\bibitem [{\citenamefont {Parker}\ and\ \citenamefont {Toms}(2009)}]{book:Parker}%
  \BibitemOpen
  \bibfield  {author} {\bibinfo {author} {\bibfnamefont {L.~E.}\ \bibnamefont {Parker}}\ and\ \bibinfo {author} {\bibfnamefont {D.~J.}\ \bibnamefont {Toms}},\ }\href@noop {} {\emph {\bibinfo {title} {Quantum Field Theory in Curved Spacetime}}}\ (\bibinfo  {publisher} {Cambridge University Press},\ \bibinfo {address} {Cambridge},\ \bibinfo {year} {2009})\BibitemShut {NoStop}%
\bibitem [{\citenamefont {Kiefer}\ \emph {et~al.}(2018)\citenamefont {Kiefer}, \citenamefont {Kwidzinski},\ and\ \citenamefont {Piechocki}}]{paper:Nick}%
  \BibitemOpen
  \bibfield  {author} {\bibinfo {author} {\bibfnamefont {C.}~\bibnamefont {Kiefer}}, \bibinfo {author} {\bibfnamefont {N.}~\bibnamefont {Kwidzinski}}, \ and\ \bibinfo {author} {\bibfnamefont {W.}~\bibnamefont {Piechocki}},\ }\href@noop {} {\bibfield  {journal} {\bibinfo  {journal} {Eur. Phys. J. C}\ }\textbf {\bibinfo {volume} {78}},\ \bibinfo {pages} {691} (\bibinfo {year} {2018})}\BibitemShut {NoStop}%
\bibitem [{\citenamefont {Parker}(1971)}]{paper:ParkerS}%
  \BibitemOpen
  \bibfield  {author} {\bibinfo {author} {\bibfnamefont {L.}~\bibnamefont {Parker}},\ }\href@noop {} {\bibfield  {journal} {\bibinfo  {journal} {Phys. Rev. D}\ }\textbf {\bibinfo {volume} {3}},\ \bibinfo {pages} {346} (\bibinfo {year} {1971})}\BibitemShut {NoStop}%
\bibitem [{\citenamefont {Srednicki}(2007)}]{book:Srednicki}%
  \BibitemOpen
  \bibfield  {author} {\bibinfo {author} {\bibfnamefont {M.}~\bibnamefont {Srednicki}},\ }\href@noop {} {\emph {\bibinfo {title} {Quantum Field Theory}}},\ \bibinfo {edition} {1st}\ ed.\ (\bibinfo  {publisher} {Cambridge University Press},\ \bibinfo {address} {Cambridge},\ \bibinfo {year} {2007})\BibitemShut {NoStop}%
\bibitem [{\citenamefont {Kolb}\ and\ \citenamefont {Turner}(1990)}]{book:Kolb}%
  \BibitemOpen
  \bibfield  {author} {\bibinfo {author} {\bibfnamefont {E.~W.}\ \bibnamefont {Kolb}}\ and\ \bibinfo {author} {\bibfnamefont {M.~S.}\ \bibnamefont {Turner}},\ }\href@noop {} {\emph {\bibinfo {title} {The Early Universe}}}\ (\bibinfo  {publisher} {Addison-Wesley},\ \bibinfo {address} {Reading, MA},\ \bibinfo {year} {1990})\BibitemShut {NoStop}%
\bibitem [{\citenamefont {Gel'fand}\ \emph {et~al.}(1963)\citenamefont {Gel'fand}, \citenamefont {Minlos},\ and\ \citenamefont {Shapiro}}]{book:Gel}%
  \BibitemOpen
  \bibfield  {author} {\bibinfo {author} {\bibfnamefont {I.~M.}\ \bibnamefont {Gel'fand}}, \bibinfo {author} {\bibfnamefont {R.~A.}\ \bibnamefont {Minlos}}, \ and\ \bibinfo {author} {\bibfnamefont {Z.~Y.}\ \bibnamefont {Shapiro}},\ }\href@noop {} {\emph {\bibinfo {title} {Representations of the rotation and Lorentz groups and their applications}}}\ (\bibinfo  {publisher} {Pergamon Press},\ \bibinfo {address} {New York},\ \bibinfo {year} {1963})\ \bibinfo {note} {translated from \textit{Predstavleniya gruppy vrashchenii i gruppy Lorentsa}, originally published by Fizmatgiz, Moscow, 1958}\BibitemShut {NoStop}%
\bibitem [{\citenamefont {Kemble}(1937)}]{book:Kemble}%
  \BibitemOpen
  \bibfield  {author} {\bibinfo {author} {\bibfnamefont {E.~C.}\ \bibnamefont {Kemble}},\ }\href@noop {} {\emph {\bibinfo {title} {The fundamental principles of quantum mechanics}}}\ (\bibinfo  {publisher} {McGraw-Hill},\ \bibinfo {address} {New York and London},\ \bibinfo {year} {1937})\BibitemShut {NoStop}%
\bibitem [{\citenamefont {Wigner}(1959)}]{book:Wigner}%
  \BibitemOpen
  \bibfield  {author} {\bibinfo {author} {\bibfnamefont {E.~P.}\ \bibnamefont {Wigner}},\ }\href@noop {} {\emph {\bibinfo {title} {Group Theory and its Application to the Quantum Mechanics of Atomic Spectra}}}\ (\bibinfo  {publisher} {Academic Press},\ \bibinfo {address} {New York},\ \bibinfo {year} {1959})\BibitemShut {NoStop}%
\bibitem [{\citenamefont {Goldberg}\ \emph {et~al.}(1967)\citenamefont {Goldberg}, \citenamefont {Macfarlane}, \citenamefont {Newman}, \citenamefont {Rohrlich},\ and\ \citenamefont {Sudarshan}}]{paper:Goldberg}%
  \BibitemOpen
  \bibfield  {author} {\bibinfo {author} {\bibfnamefont {J.~N.}\ \bibnamefont {Goldberg}}, \bibinfo {author} {\bibfnamefont {A.~J.}\ \bibnamefont {Macfarlane}}, \bibinfo {author} {\bibfnamefont {E.~T.}\ \bibnamefont {Newman}}, \bibinfo {author} {\bibfnamefont {F.}~\bibnamefont {Rohrlich}}, \ and\ \bibinfo {author} {\bibfnamefont {E.~C.~G.}\ \bibnamefont {Sudarshan}},\ }\href@noop {} {\bibfield  {journal} {\bibinfo  {journal} {J. Math. Phys.}\ }\textbf {\bibinfo {volume} {8}},\ \bibinfo {pages} {2155} (\bibinfo {year} {1967})}\BibitemShut {NoStop}%
\bibitem [{\citenamefont {Sakurai}(1993)}]{book:Sakurai}%
  \BibitemOpen
  \bibfield  {author} {\bibinfo {author} {\bibfnamefont {J.~J.}\ \bibnamefont {Sakurai}},\ }\href@noop {} {\emph {\bibinfo {title} {Modern Quantum Mechanics}}},\ \bibinfo {edition} {revised ed.}\ ed.\ (\bibinfo  {publisher} {Addison-Wesley},\ \bibinfo {address} {Boston},\ \bibinfo {year} {1993})\BibitemShut {NoStop}%
\bibitem [{\citenamefont {Edmonds}(1957)}]{book:Ed}%
  \BibitemOpen
  \bibfield  {author} {\bibinfo {author} {\bibfnamefont {A.~R.}\ \bibnamefont {Edmonds}},\ }\href@noop {} {\emph {\bibinfo {title} {Angular Momentum in Quantum Mechanics}}}\ (\bibinfo  {publisher} {Princeton University Press},\ \bibinfo {address} {Princeton, NJ},\ \bibinfo {year} {1957})\BibitemShut {NoStop}%
\bibitem [{\citenamefont {Davydov}(1965)}]{book:Dav}%
  \BibitemOpen
  \bibfield  {author} {\bibinfo {author} {\bibfnamefont {A.~S.}\ \bibnamefont {Davydov}},\ }\href@noop {} {\emph {\bibinfo {title} {Quantum Mechanics}}}\ (\bibinfo  {publisher} {Pergamon Press},\ \bibinfo {address} {Oxford},\ \bibinfo {year} {1965})\BibitemShut {NoStop}%
\bibitem [{\citenamefont {Winter}(1954)}]{paper:Winter}%
  \BibitemOpen
  \bibfield  {author} {\bibinfo {author} {\bibfnamefont {C.~V.}\ \bibnamefont {Winter}},\ }\href@noop {} {\bibfield  {journal} {\bibinfo  {journal} {Physica}\ }\textbf {\bibinfo {volume} {20}},\ \bibinfo {pages} {274} (\bibinfo {year} {1954})}\BibitemShut {NoStop}%
\bibitem [{\citenamefont {King}\ \emph {et~al.}(1943)\citenamefont {King}, \citenamefont {Hainer},\ and\ \citenamefont {Crossl}}]{paper:CKing}%
  \BibitemOpen
  \bibfield  {author} {\bibinfo {author} {\bibfnamefont {G.~W.}\ \bibnamefont {King}}, \bibinfo {author} {\bibfnamefont {R.~M.}\ \bibnamefont {Hainer}}, \ and\ \bibinfo {author} {\bibfnamefont {P.~C.}\ \bibnamefont {Crossl}},\ }\href@noop {} {\bibfield  {journal} {\bibinfo  {journal} {J. Chem. Phys.}\ }\textbf {\bibinfo {volume} {11}},\ \bibinfo {pages} {27} (\bibinfo {year} {1943})}\BibitemShut {NoStop}%
\bibitem [{\citenamefont {Mulliken}(1941)}]{paper:Mull}%
  \BibitemOpen
  \bibfield  {author} {\bibinfo {author} {\bibfnamefont {R.~S.}\ \bibnamefont {Mulliken}},\ }\href@noop {} {\bibfield  {journal} {\bibinfo  {journal} {Phys. Rev.}\ }\textbf {\bibinfo {volume} {59}},\ \bibinfo {pages} {873} (\bibinfo {year} {1941})}\BibitemShut {NoStop}%
\bibitem [{\citenamefont {Landau}\ and\ \citenamefont {Lifshitz}(1981)}]{book:Landau3}%
  \BibitemOpen
  \bibfield  {author} {\bibinfo {author} {\bibfnamefont {L.~D.}\ \bibnamefont {Landau}}\ and\ \bibinfo {author} {\bibfnamefont {E.~M.}\ \bibnamefont {Lifshitz}},\ }\href@noop {} {\emph {\bibinfo {title} {Quantum Mechanics: Non-Relativistic Theory}}},\ \bibinfo {edition} {3rd}\ ed.,\ \bibinfo {series} {Course of Theoretical Physics}, Vol.~\bibinfo {volume} {3}\ (\bibinfo  {publisher} {Butterworth-Heinemann},\ \bibinfo {address} {Oxford},\ \bibinfo {year} {1981})\BibitemShut {NoStop}%
\bibitem [{\citenamefont {Vleck}(1929)}]{paper:VV}%
  \BibitemOpen
  \bibfield  {author} {\bibinfo {author} {\bibfnamefont {J.~H.~V.}\ \bibnamefont {Vleck}},\ }\href@noop {} {\bibfield  {journal} {\bibinfo  {journal} {Phys. Rev.}\ }\textbf {\bibinfo {volume} {33}},\ \bibinfo {pages} {467} (\bibinfo {year} {1929})}\BibitemShut {NoStop}%
\end{thebibliography}%

 \end{document}